\documentclass{aastex63}

\usepackage{amsmath}
\usepackage{multirow}

\submitjournal{Astronomy Reports}

\shorttitle{Inverse Compton Scattering of a Central Source Radiation as X-rays mechanisms}
\shortauthors{Butuzova et al.}

\begin{document}

\title{Inverse Compton Scattering of a Central Source Radiation as a Possible Mechanism for \\the Formation of X-Ray Radiation from Kiloparsec Jets of Core-Dominated Quasars}

\correspondingauthor{Marina S. Butuzova}
\email{mbutuzova@craocrimea.ru}

\author[0000-0001-7307-2193]{Marina S. Butuzova}
\affiliation{Crimean Astrophysical Observatory of RAS, Nauchny, 298409, Russia}

\author[0000-0002-9702-2307]{Alexander B. Pushkarev}
\affiliation{Crimean Astrophysical Observatory of RAS, Nauchny, 298409, Russia}
\affiliation{Astro Space Center of Lebedev Physical Institute, Profsoyuznaya 84/32, Moscow 117997, Russia}

\author[0000-0003-2914-2507]{Elena S. Shablovinskaya}
\affiliation{Special Astrophysical Observatory of RAS,  Nizhny Arkhyz, 369167, Russia}

\author{Sergey V. Nazarov}
\affiliation{Crimean Astrophysical Observatory of RAS, Nauchny, 298409, Russia}

\begin{abstract}

For the interpretation of X-ray radiation from kiloparsec jets of quasars, the inverse Compton scattering
of the cosmic microwave background has been widely used for almost 20 years. A recent analysis of the
\textit{Fermi}-LAT observational data showed that this assumption is inapplicable for jets of several quasars. In this
paper, we consider the inverse Compton scattering of photons from a central source as a possible mechanism
for the formation of X-ray radiation from kiloparsec jets of the quasars 
PKS~0637$-$752, 3C~273, PKS~1510$-$089, and PKS~1045$-$188.
Estimates for the angle between the line of sight and the velocity of kiloparsec-scale jets
are obtained. The predicted gamma-ray flux for all objects turned out to be below the upper limit on the flux
from a kiloparsec jet obtained from the \textit{Fermi}-LAT data. It is shown that our assumption about the mechanism
of kiloparsec jets X-ray radiation is consistent with all data of multiwavelength observations available to date.
\end{abstract}


\section{Introduction} \label{sec:intro}

Kiloparsec (kpc) jets of active galactic nuclei have
been observed with a high angular resolution in the
X-ray range by the \textit{Chandra} space observatory since
2000 \citep{Schwartz2000}.
For the first discovered jet of the quasar PKS~0637–752, as well as for jets of other core-dominated
quasars, the X-ray flux turned out to be higher than
expected from the extrapolation of the synchrotron
radio–optical spectrum to X-ray frequencies \citep{Schwartz2000,Harris06}.
This is an evidence of different processes that generate
radiation in these ranges. In one of the first works, it
was shown for the PKS~0637$-$752 quasar jet that the
most probable mechanism of its high-frequency radiation
is inverse Compton scattering (IC) of self synchrotron
radiation \citep{Schwartz2000}. However, from a comparison
of the radio and X-ray fluxes, it followed
that the condition of the energy equipartition between the magnetic field and the particles is not fulfilled: a lager energy is contained in the emitting particles. \citet{Tav00,Cel01} suggested that the X-ray radiation of the PKS~0637$-$752 jet is formed due to inverse Compton scattering of cosmic microwave background (IC/CMB). It was assumed that a kpc-scale jet, by analogy with a parsec (pc) jet \citep{Chartas00}, moves as a whole with an ultrarelativistic velocity at a small angle to the line of sight.
In this case, the energy density of the cosmic microwave background (CMB)
in the reference frame of the kpc-scale jet increases, which leads to an increase in the X-ray flux, while the flux at radio frequencies being the same, and, consequently, to the fulfillment of the energy equipartition condition. This model of the formation of X-ray radiation, named ``beamed IC/CMB'', later became widely used for the interpretation of X-ray
radiation from kpc-scale jets of core-dominated quasars \citep[see, e.g., ][]{Samb04,Marsh05,HoganList11,Marsh11,Marsh18}.

Recent studies have shown that the beamed IC/CMB model is inapplicable to high-frequency radiation from the kpc-scale jets of the quasars PKS~0637$-$752 \citep{Meyer15} and 3C 273 \citep{MeyGeor14}, because, in the observational
data of \textit{Fermi}-LAT in the gamma-ray range, there is no high constant flux characterized by a hard spectrum, which must have been generated in kpc-scale jets. An indirect argument against the beamed IC/CMB model is the fact that there is no statistically significant difference in the distribution
of the difference in the positional angles of the pc- and kpc-scale jets of core-dominated quasars that have (27 objects) or do not have (23 objects) a
detectable X-ray flux from a kpc-scale jet \citep{But16}.

Alternative mechanisms for the formation of X-ray radiation from kpc-scale jets of quasars, e.g., such as synchrotron radiation, produced either by the second high-energy population of electrons \citep{DA04} or by protons \citep{Ahar02}, introduce additional free parameters, by varying which one can obtain acceptable estimates of the physical conditions in the kpc-scale jets. In addition, these mechanisms require two different acceleration mechanisms to operate in one part (or in closely located parts) of the jet, which complicates the consideration of the jet physical nature.

On the other hand, if we dispense with the a priori small angle ($<10^\circ$) between the kpc-scale jet and the line of sight, then the kpc-scale jets observed in projection are actually at a smaller distance from the active nucleus. Then, at least for the knots of kpc-scale jets nearest to the nucleus, X-ray radiation can be formed due to inverse Compton scattering of the radiation from the central source (IC/CS).
By the term ``knot'' we mean a part in a jet with an increased surface
brightness, which is probably caused by an increased density of emitting particles. Another reason for which IC/CS is becoming a more attractive mechanism for the formation of X-ray radiation is that the
CS radiation at frequencies from radio to millimeter
wavelengths is generated in a pc-scale jet \citep{Kovalev2005}
and, due to relativistic effects, is amplified in the
frame of reference of a kpc-scale jet so that IC/CS
contributes to the observed X-ray radiation significantly
more than IC/CMB.
As was shown for kpc-scale
jets of the quasars 3C~273 \citep{MBK10} and PKS~1127$-$145
\citep{BP19}, IC/CS provides a natural explanation for the
observed decrease with distance in the X-ray intensity
of knots from the active nucleus and allows one, with
a known CS spectrum, to determine both the physical
parameters of the knots and the angle between the line
of sight and the velocity of the kpc-scale jet \cite{BP19}. Since
a decrease in the X-ray intensity is often observed
along the kpc-scale jets of quasars \citep{Harris06,Samb04}, this may
indicate the widespread occurrence of IC/CS. Previously,
this mechanism was considered only for radio
galaxies \citep{Cel01,StawOB08, Ostor10,Migliori12}.

In this paper, we investigate the applicability of
IC/CS to the interpretation of X-ray radiation from
kpc-scale jets of quasars. In Section 2, we explain our
choice of objects and present the available observational
data, which we use to determine the angle
between the line of sight and the velocity of kpc-scale
jets (Section 3). In Section 4, we estimate the magnetic
field strength and the density of emitting electrons.
The emission spectrum of jets in the X-ray and
gamma-ray ranges is simulated for comparison with
the upper bound of the flux from the kpc-scale jets in
the gamma-ray range. A discussion of the results and
conclusions is presented in Section 5.

\section{Objects and method of research}

The selection of objects was made according to the
following criteria. First, only core-dominated quasars
were selected. Second, a decrease in the X-ray intensity
of the knots along the kpc-scale jet, which is an
indication of the possibility of IC/CS radiation,
should be observed. In this case, jets with two or more
knots observed both in the radio and in the X-ray
ranges were selected.
Third, it was necessary to have
observational data of pc-scale jets, from which it is
possible to estimate the apparent superluminal velocity
of the components and, consequently, to estimate
the velocity $\beta$ (in units of the speed of light $c$) and the
angle of the pc-scale jet with the line of sight $\theta_{text{pc}}$. To
simplify the selection of objects, we used web pages
containing data on X-ray jets\footnote{http://hea-www.harvard.edu/XJET/} and observational data from the MOJAVE program\footnote{http://www.physics.purdue.edu/astro/MOJAVE/allsources.html}.

These criteria are met by the following objects: 3C~273, PKS~1127$-$145, PKS~1045$-$188, PKS~1510$-$089. In the framework of IC/CS the angle between the line of sight and the kpc-scale jet of the quasar 3C~273 was determined by \cite{MBK10} from comparison of the integral CS and CMB energy densities rather than by comparing the X-ray flux produced in IC/CS and IC/CMB.
Therefore, 3C~273 was left for this study. The geometrical and kinematic parameters of the kpc-scale jet of the quasar PKS~1127$-$145 under the assumption of IC/CS were determined by \citet{BP19} and will be used here for comparison.
Despite the fact that the X-ray intensity of the knots of the quasar PKS~0637$-$752 kpc-scale jet does not exhibit a distinct decrease with the
distance from the core, we included this object in the sample: it is interesting as a prototype for which the beamed IC/CMB scenario was first suggested and then rejected.

\begin{figure}
    \centering
    \includegraphics[scale=0.8]{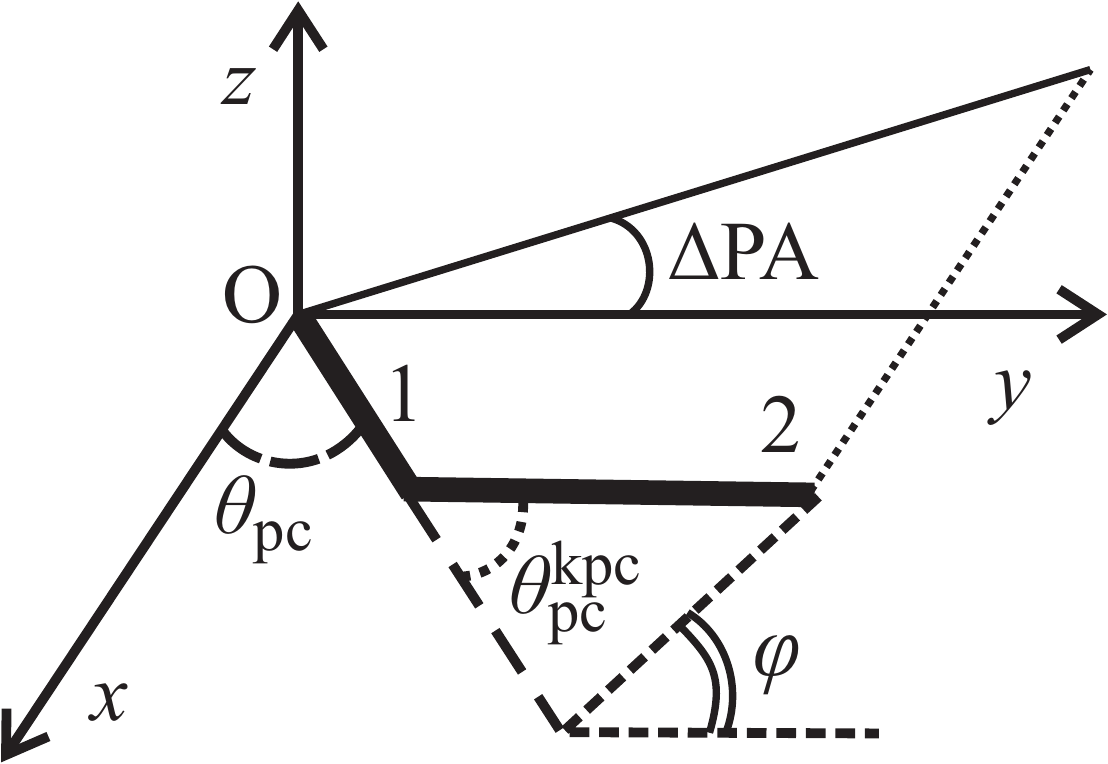}
    \caption{Schematic diagram of the jet bend between the (1) pc- and (2) kpc-scales. Axis $x$ is directed along the line of sight, and axis $y$ is towards along the projection of the pc-scale jet onto the plane of the sky. The angles, used in expression~(\ref{eq:th-pc-kpc}) to define the jet bend angle $\theta_{\text{kpc}}^{\text{pc}}$, are denoted.}
    \label{fig:izgib}
\end{figure}

\begin{figure}
    \centering
    \includegraphics[scale=0.8]{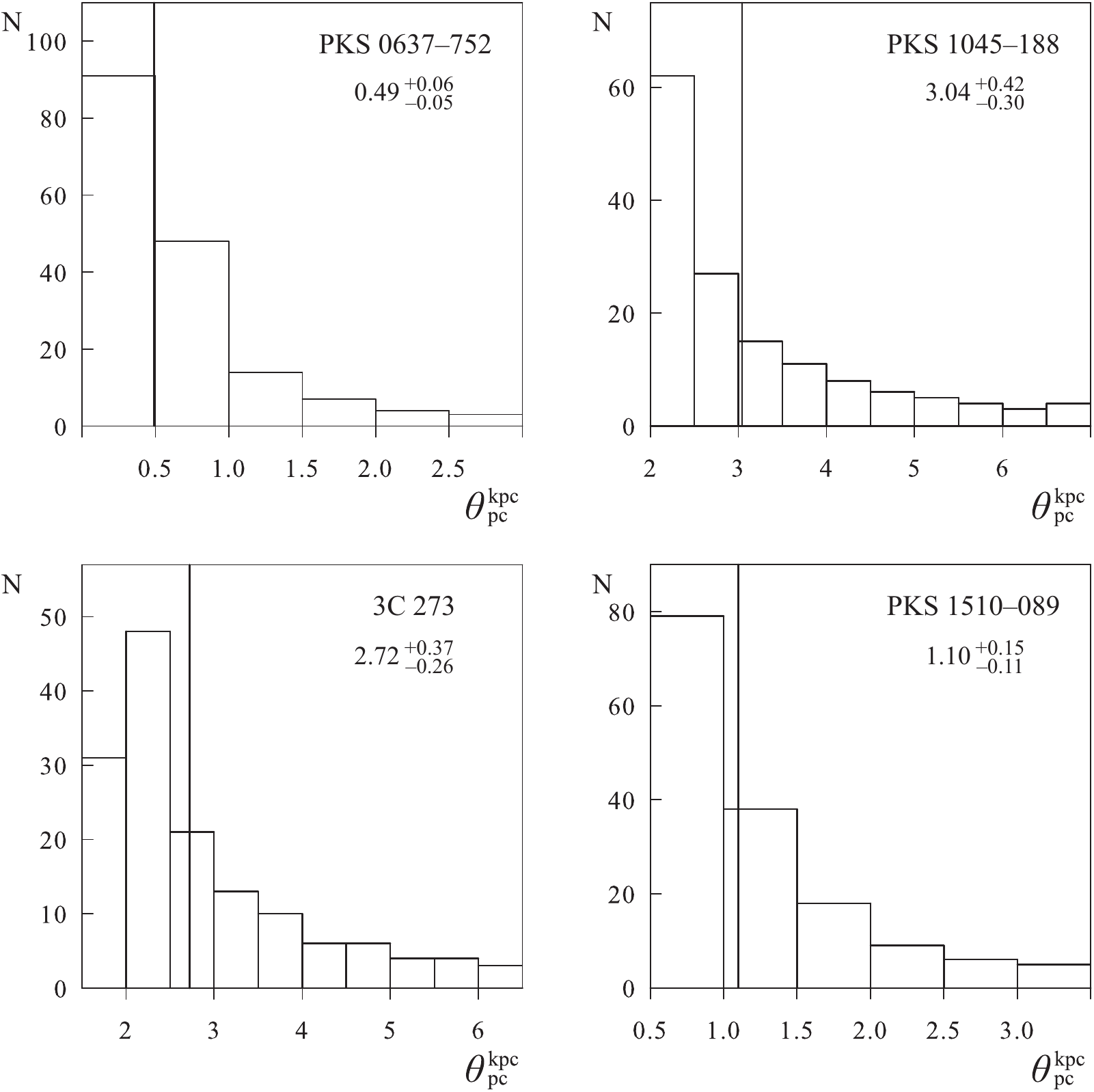}
    \caption{Distribution of $\theta_{\text{kpc}}^{\text{pc}}$ as $\varphi$ varies within acceptable limits with a step of $1^\circ$. In the case of PKS~1045$-$188, 3C~273, and PKS~1510$-$089, there are no values of $\theta_{\text{kpc}}^{\text{pc}}$ smaller than the lower bound of the given intervals. The graphs were plotted using at least $86\%$ of the $\theta_{\text{kpc}}^{\text{pc}}$
 values obtained for various $\varphi$. The solid vertical line represents the median values of $\theta_{\text{kpc}}^{\text{pc}}$ within the presented range. The median value of $\theta_{\text{kpc}}^{\text{pc}}$ is given below the object's name.}
    \label{fig:izgibvalue}
\end{figure}

In this study, we use formulae that were derived and thoroughly described by \citet{BP19}. Here, we will emphasize only the main ones. To take into account the spectra of both relativistic electrons and scattered radiation, the
expression for the scattered radiation flux density is found using the invariant kinetic Boltzmann equation for IC \citep{NagirnerP94,NagirnerP01,Nagirner94}.
 Since the dominant part of the scattered radiation is formed in a relativistic pc-scale jet \citep{Kovalev2005}, in order to find the distribution function of scattered photons in the reference frame of the kpc-scale jet from the observed spectrum of the quasar, appropriate transformations should be made.
The transition to the reference frame of the pc-scale jet was carried out using the Doppler factor $\delta=\left(1-\beta^2_{\text{pc}}\right)^{1/2}/\left(1-\beta_{\text{pc}}\cos \theta_{\text{pc}}\right)$.  
The velocity $\beta_{\text{pc}}$ and viewing angle $\theta_{\text{pc}}$ of a pc-scale jet is determined from the apparent velocity  
$\beta_{\text{app}}$ of its features from the expressions $\theta_{\text{pc}}=\left(1+\beta_{\text{app}}^2 \right)^{-0.5}$ and  $\theta_{\text{pc}}\sim 1/\Gamma_{\text{pc}}$ (where $\Gamma_{\text{pc}}=\left(1-\beta^2_{\text{pc}}\right)^{-1/2}$ --- is the Lorentz factor). For the transition to the pc-scale jet reference frame, we use the Doppler factor of the pc-scale jet $\delta_{\text{j}}$, which would be observed from the pc-scale jet.    
The value $\delta_{\text{j}}$ depends on the angle $\theta^\prime\,_{\text{pc}}^{\text{kpc}}$ between the velocity vector of the pc-scale jet and the direction from the pc- to the kpc-scale jet (the prime denotes the values in the reference frame of the kpc-scale jet).
In other words, $\theta^\prime\,_{\text{pc}}^{\text{kpc}}$ is the angle of the actual jet bend, which occurs between the pc- and kpc-scales. The bend angle $\theta_{\text{pc}}^{\text{kpc}}$ in the observer's reference frame can be found from the difference in the positional angles of the pc- and kpc-scale jets, $\Delta{\text{PA}}$, using the formula~(1) in \citep{CM93}\footnote{$\Delta{\text{PA}}$, $\theta_{\text{pc}}$, $\theta_{\text{pc}}^{\text{kpc}}$ in \citet{CM93} are denoted as $\eta$, $\theta$, $\zeta$, espectively}:
\begin{equation}
\cot \theta_{\text{pc}}^{\text{kpc}}=\frac{\sin\varphi - \tan\Delta{\text{PA}}\cos\theta_{\text{pc}}\cos\varphi}{\tan\Delta{\text{PA}}\sin\theta_{\text{pc}}},
\label{eq:th-pc-kpc}
\end{equation}
where $\varphi$ is the azimuth bend angle (Fig.~\ref{fig:izgib}). Figure~\ref{fig:izgib} shows a scheme of the jet bend between the pc- and kpc-scales and the direction of reference $\varphi$ is determined. It can be seen that, if the pc-scale jet deviates from the pc-scale jet in the direction of the $z$ axis, then $0^\circ<\varphi<180^\circ$, as is the case of PKS~0637$-$752 and PKS~1045$-$188. For each $\varphi$ in the interval specified for the object under consideration
and changing with a step of $1^\circ$, we obtained the values of $\theta\,_{\text{pc}}^{\text{kpc}}$ from formula~(\ref{eq:th-pc-kpc}).
Under this, the initial value of $\varphi$ was greater than the lower bound of the considered interval of possible values of $\varphi$ by $0.1^\circ$. The greatest values (of several dozen degrees) correspond to $\varphi\approx\Delta{\text{PA}}$ ($\varphi\approx\Delta{\text{PA}}+180^\circ$ for PKS~1510$-$089). In the vast majority of cases, $\theta\,_{\text{pc}}^{\text{kpc}}<10^\circ$.
Figure~\ref{fig:izgibvalue} shows histograms of $\theta_{\text{kpc}}^{\text{pc}}$ in the intervals containing at least $86\%$ of values. The distributions of $\theta_{\text{kpc}}^{\text{pc}}$ are not described by standard statistical distributions; therefore, as an estimate of $\theta_{\text{kpc}}^{\text{pc}}$, we used the median value rounded to degrees from those in the intervals presented in Fig.~\ref{fig:izgibvalue}.
The value of $\theta_{\text{kpc}}^{\text{pc}}$ was obtained for each object is presented in Table~\ref{tab:info}. Note that formula~(\ref{eq:th-pc-kpc}) does not take into account the relativistic
aberration, which, when the jet decelerates, will lead to an increase in the angle between the jet and the line of sight.
This change will occur in the plane orthogonal to the plane of the sky and containing the jet axis
and the line of sight. If the kpc-scale jet has a velocity $\beta_{\text{pc}}$, then the angle used in the calculation of $\delta_{\text{j}}$ is determined by the formula
\begin{equation}
\theta^\prime\,_{\text{pc}}^{\text{kpc}}=\arctan\left(\frac{\beta_{\text{pc}}\sin\theta_{\text{pc}}^{\text{kpc}}\sqrt{1-\beta_{\text{kpc}}^2}}{\beta_{\text{pc}}\cos\theta_{\text{pc}}^{\text{kpc}}-\beta_{\text{kpc}}} \right).
\label{eq:th-pc-kpc-sh}
\end{equation}

In following estimates, we use expression~(\ref{eq:th-pc-kpc}), because there are no a priori values for the velocity and the viewing angle of the kpc-scale jet.
When the parameters $\theta_{\text{kpc}}$ and $\beta_{\text{kpc}}$ will be estimated within IC/CS, it will be possible to use formula~(\ref{eq:th-pc-kpc-sh}), which ultimately will lead to a change in the estimates of $\theta_{\text{kpc}}$ and $\beta_{\text{kpc}}$.
After several iterations, it is possible to achieve well-matched values of all considered parameters.
But, as \citet{BP19} noted, with a several-fold increase in the angle $\theta_{\text{kpc}}^{\text{pc}}$, the estimates of the number density of emitting electrons and the magnetic field strength do not change significantly.
As will be shown in Section~3, $\theta_{\text{kpc}}$ can be at best determined in a certain interval, but,for a part of sources, only the lower bound of $\theta_{\text{kpc}}$ can be obtained.
Therefore, the use of only the value of $\theta_{\text{kpc}}^{\text{pc}}$
 obtained from expression~(\ref{eq:th-pc-kpc}) in further calculations is justified.
It should also be noted that there is observational evidence that kpc-scale jets have a moderately relativistic propagation speed \citep{Homan15, WA97,ArLon04,MullinH09} and the smaller $\beta_{\text{kpc}}$, the smaller the difference between $\theta\,_{\text{pc}}^{\text{kpc}}$ and  $\theta^\prime\,_{\text{pc}}^{\text{kpc}}$.

\begin{figure}
    \centering
    \includegraphics[scale=0.8]{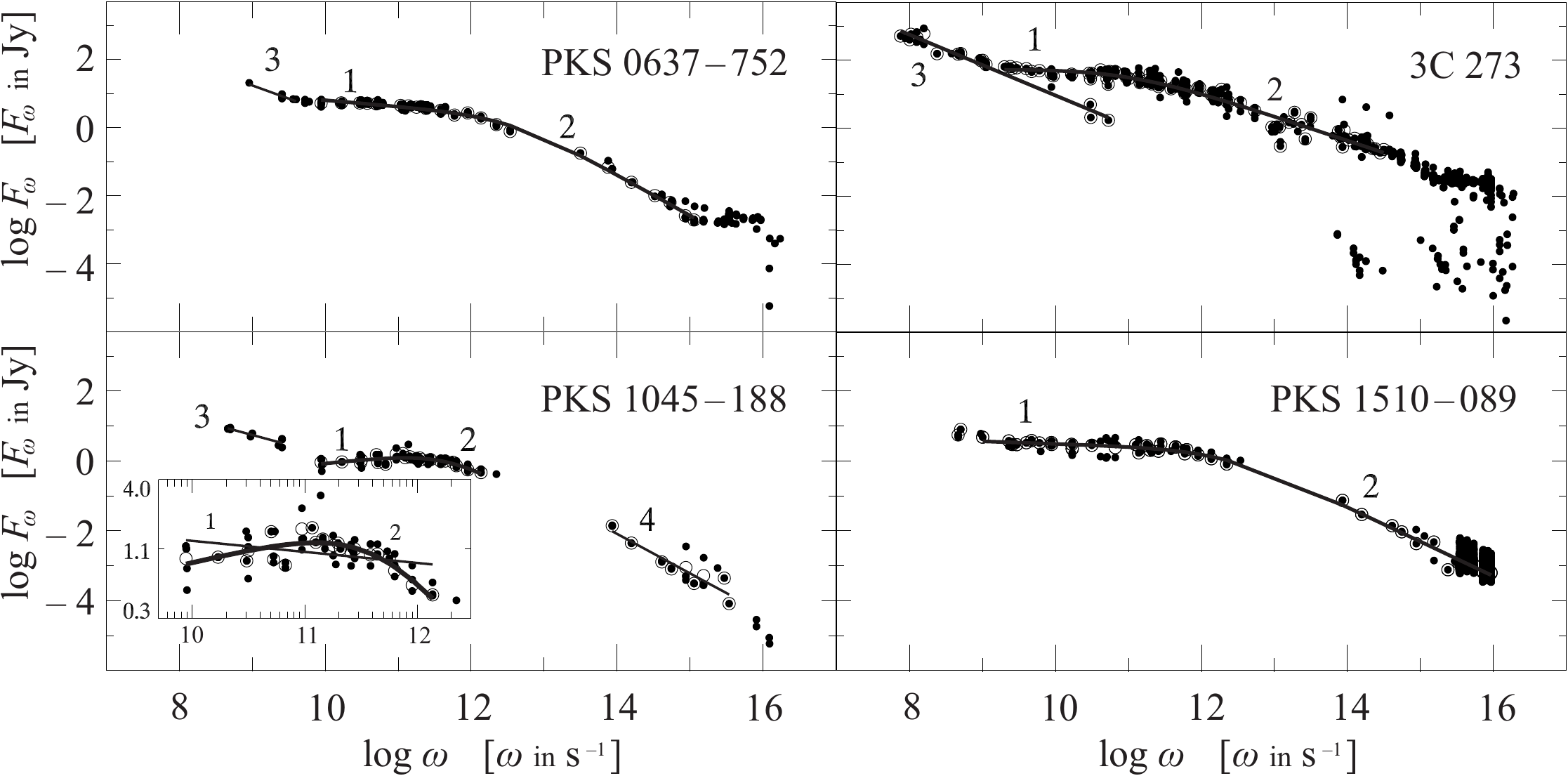}
    \caption{Observed spectra of quasars and their approximations. The circles represent flux-averaged observations at one frequency.
For the quasar PKS~1045$-$089, the radio spectrum is additionally shown in close-up to display approximations by both one and two power-laws.}
    \label{fig:CSspectra}
\end{figure}

To determine the distribution function of scattered
photons, it is assumed that the emission spectrum is power-law (radiation flux density is $F_\nu=Q\,\omega^{-\alpha}$, where $\alpha$ is the spectral index).
The spectra of the objects under consideration in the range from radio to millimeter wavelengths are shown in Fig.~\ref{fig:CSspectra}, from which it can be seen that they can be described by several power-law parts.
The indices``1'' and ``2'' denote the parts of the CS's radiation spectrum that we associate with the radiation of a pc-scale jet generated in optically
thick and thin media, respectively. Index ``3'' marks the parts, which we associate with the low-frequency radiation of the kpc-scale jet. The model parameters of the CS's radiation spectra are given in Table~\ref{tab:CSspectra}. For a more detailed description of the turn between parts 1 and 2, we used a function of the form  $F_\nu\propto\left[b\left(\omega/\omega_0 \right)^{-\alpha_1}+d\left( \omega/\omega_0\right)^{-\alpha_2} \right]^{-1}$, where $b$ and $d$ are
parameters; $\alpha_{1}$ and $\alpha_2$ are the spectral indices of parts 1 and 2, respectively; and $\omega_0$ is the turn-over frequency of the spectrum between parts 1 and 2.
We took the observational data for each of the objects from the NED database\footnote{http://ned.ipac.caltech.edu/}; they are mainly presented in \citep{Courvoisier98, Turler99} for 3C~273, \citep{Kovalev2005, Geld1981, Wright1990, Slee1995, Adraou2001, Mingaliev2015, Wright2009, Vollmer2010} for PKS~1045$-$188, \citep{Slee1995, Gear94, Hovatta08} for PKS~1510$-$089, \citep{Moshir1990, EdwardsPT06, Wright2009, Tingay03, Burgess2006} for PKS~0637$-$752.

\begin{table*}
 \caption{
Basic information on the analyzed quasars and their parsec jets. 
The columns represent: (1) object; (2) redshift; (3) luminosity distance (in this article, we use the $\Lambda$CDM-model with the following parameters: $H_0=71$~km~s$^{-1}$~Mpc$^{-1}$, $\Omega_{\text{m}}=0.27$, and $\Omega_\lambda=0.73$ \citep{Komatsu09}); (4) correspondence between angular and linear scales; (5) the apparent average velocity of the pc-scale jet features; (6), (7) the velocity and the angle with the line of sight of the pc-scale jet, respectively, estimated from $\beta_{\text{app}}$; (8), (9) positional angles of pc- and kpc-scale jets, respectively; (10) the difference between the positional angles of pc- and kpc-scale jets; (11) the actual bend angle of the jet between the pc- and kpc-scales in the observer’s reference frame. 
The data in columns (7)$-$(11) are in degrees.
 } 
 \label{tab:info} 
  \begin{flushleft}
  \begin{tabular}
 {|c|c|c|c|c|c|c|c|c|c|c|}
  \hline
 Object & $z$ & $D_L$, Mpc & kpc in 1$^{\prime\prime}$ &
 $\beta_\text{app}$ & $\beta_\text{pc}$ & $\theta_\text{pc}$, $^\circ$ & PA$_\text{pc}$, $^\circ$  &  PA$_\text{kpc}$, $^\circ$ & $\Delta$PA, $^\circ$ & $\theta_\text{pc}^\text{kpc}$, $^\circ$ \\
  \hline
 (1) & (2) & (3) & (4) & (5) & (6) & (7) & (8) & (9) & (10) & (11) \\
  \hline
    PKS~0637$-$752 & 0.653 & 3909.3 & 6.94 & $13.3\pm1.0$ \footnote{\citet{EdwardsPT06}} & 0.997 & 4 & 273 $^{\text{a}}$ & 278 \footnote{\citet{Schwartz2000}} & 5 & 1 \\
  PKS~1045$-$188 & 0.595 & 3487.7 & 6.65 & $10.35\pm0.32$ \footnote{\citet{Lister19}} & 0.995 & 6 & 146 \footnote{\citet{KharbLC10}} & 125 $^{\text{c}}$ & 21 & 3\\
  3C~273 & 0.158 & 747.0 & 2.70 & $8.09\pm 0.06$ $^{\text{b}}$ & 0.992 & 7 & 238 $^{\text{c}}$ & 222 $^{\text{c}}$ & 16 & 3 \\
  PKS~1510$-$089 & 0.36 & 1907.0 & 5.00 & $18.4\pm2.4$ $^{\text{b}}$ & 0.999 & 3 & 328 $^{\text{c}}$ & 163 $^{\text{c}}$ & 165 & 1\\
  \hline
  \end{tabular}
 \end{flushleft}
\end{table*}

\begin{table}
 \caption{
Parameters of approximation of the spectra of central sources by power-laws 
 } 
 \label{tab:CSspectra} 
 \begin{flushleft}
 \begin{tabular}
 {|c|c|c|c|c|c|}
  \hline
 Object & Part & Range $\omega$, s$^{-1}$ & $Q$, $10^{-23}$ erg/sm$^2$/s$^\alpha$ & $\alpha$ & $\chi^2$ \\
  \hline
 \multirow{3}{*}{PKS 0637$-$752} &  1 & $8 \cdot 10^9 - 3.7 \cdot 10^{12}$ & $ 412.13 \pm 40.37$ & $0.18 \pm 0.04$ & \multirow{2}{*}{0.095} \\
\cline{2-5}
            &  2 & $3.7 \cdot 10^{12} - 2 \cdot 10^{15}$ & $(2.67 \pm 0.75) \cdot 10^{15}$ & $1.20 \pm 0.05$ & \\
\cline{2-6}
            &  3 & $9 \cdot 10^8 - 4 \cdot 10^{9}$ & $(1.57 \pm 0.11) \cdot 10^{8}$ & $0.77 \pm 0.10$ & 0.022 \\
 \hline          
 \multirow{5}{*}{PKS 1045$-$188} &  1 & $8 \cdot 10^9 - 4.3 \cdot 10^{11}$ & $ (6.1 \pm 4.4) \cdot 10^{-3}$ & $-0.21 \pm 0.01$ & \multirow{2}{*}{0.099} \\
\cline{2-5}
            &  2 & $4.3 \cdot 10^{11} - 2 \cdot 10^{12}$ & $(1.93 \pm 0.10) \cdot 10^{13}$ & $1.12 \pm 0.04$ & \\
\cline{2-6}
            &  3 & $4 \cdot 10^8 - 3 \cdot 10^{9}$ & $(6.30 \pm 0.19) \cdot 10^5$ & $0.56 \pm 0.08$ & 0.057 \\
\cline{2-6}
            &  1-2 & $8 \cdot 10^9 - 2 \cdot 10^{12}$ & $9.97 \pm 2.33$ & $0.13 \pm 0.05$ & 0.313 \\
\cline{2-6}
            &  4 & $8 \cdot 10^{13} - 4 \cdot 10^{15}$ & $(5.34 \pm 0.87) \cdot 10^{13}$ & $1.13 \pm 0.15$ & 0.371 \\
  \hline
 \multirow{3}{*}{3C 273} &  1 & $2 \cdot 10^9 - 9.3 \cdot 10^{10}$ & $ 25.45 \pm 3.18$ & $-0.04 \pm 0.13$ & \multirow{2}{*}{2.106} \\
\cline{2-5}
            &  2 & $9.3 \cdot 10^{10} - 3 \cdot 10^{14}$ & $(4.28 \pm 0.01) \cdot 10^{9}$ & $0.71 \pm 0.05$ & \\
\cline{2-6}
            &  3 & $7 \cdot 10^7 - 6 \cdot 10^{10}$ & $(7.89 \pm 0.19) \cdot 10^{9}$ & $0.89 \pm 0.04$ & 0.255 \\
\hline
     \multirow{2}{*}{PKS 1510$-$089} &  1 & $1 \cdot 10^9 - 2.2 \cdot 10^{12}$ & $ 15.44 \pm 0.89$ & $-0.07 \pm 0.03$ & \multirow{2}{*}{ 0.472} \\
\cline{2-5}
            &  2 & $2.2 \cdot 10^{12} - 2 \cdot 10^{16}$ & $3.66\pm0.044)\cdot10^{12}$ & $0.99 \pm 0.04$  &  \\
\hline
  \end{tabular}
   \end{flushleft}
\end{table}

\begin{figure}
\center{ \includegraphics[scale=0.6]{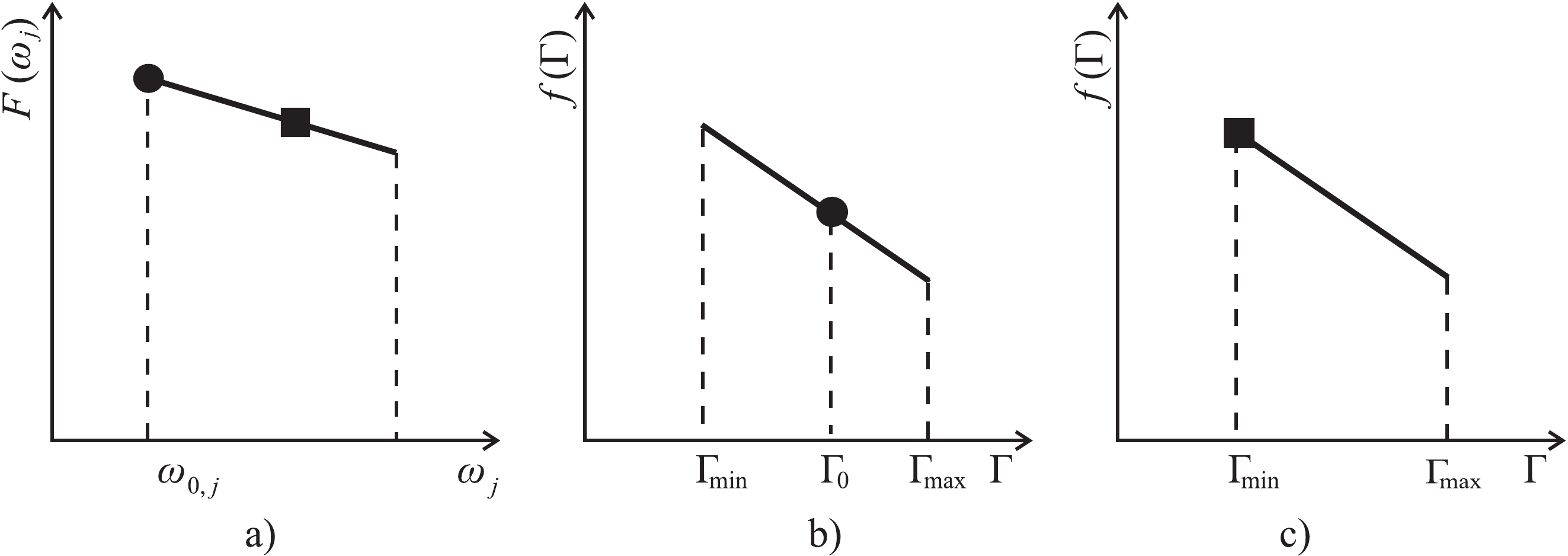}  }  
 \caption{Schematics of power-law spectra of scattered radiation  (a) and ultrarelativistic electrons (b, c). The same symbols mark
the parts of the electron and photon spectra that make the main contribution to the observed X-ray radiation under IC at the fixed frequency. The limitation imposed on IC by (b) the photon spectrum and (c) electron spectrum are illustrated.}
 \label{fig:restrinctions}
\end{figure}

 As \citet{BP19} shown, for IC of power-law photon and electron energy spectra, two cases are possible, each of which is characterized by its own spectral index of scattered radiation, $\alpha_\text{X}$.
In the first case, under the limitation from the photon spectrum, the main contribution to the scattered radiation at a given frequency is made by the IC of photons with a frequency corresponding
to one of the boundaries of the photon spectrum by electrons with energies far from the boundary
values (see Fig.~\ref{fig:restrinctions}b).
Then the spectral flux density of the scattered radiation is determined by the formula
\begin{equation}
\begin{split}
F\left(\omega_{\text{X}}\right)=\left(1+z\right)^{\alpha-\left(\gamma-1\right)/2} 
A \frac{\left[ 2\left( 1-\cos\theta_{\text{kpc}}\right) \right]^{\left(\gamma+1\right)/2} }{\left( \gamma+1\right) \left|\gamma-2\alpha-1 \right| }
\omega_{\text{cut, j}}^{\left(\gamma-1\right)/2-\alpha}\omega_{\text{X}}^{-\left(\gamma-1\right)/2},
\end{split}
\label{eq:FwxPH}
\end{equation}
where
\begin{equation*}
 A=\left(\frac{\delta_{\text{j}}}{\delta} \right)^{3+\alpha} r_e^{2}
\frac{V \sin^2\theta_{\text{kpc}}}{R^2} \left( m_e c^2\right)^{1-\gamma} \times\mathcal{K} Q, 
\end{equation*}

$z$ is the redshift of the object, $\alpha$ is the spectral index of the scattering photon spectrum, $\gamma$ is the spectral index of the electron energy distribution, $r_e$ and $m_e$ are the classical radius and rest mass of the electron, $V$ is the volume of the emission region (the knot of the kpc-scale jet), $\theta_{\text{kpc}}$ is the angle between the kpc-scale jet and the line of sight, $R$ is the distance from CS to the considered knot of the kpc-scale jet, $\mathcal{K}$ is the coefficient of proportionality of the electron energy distribution, and $\omega_{\text{X}}$ is the observed X-ray frequency. If the
spectral index of the radio radiation of the kpc-scale jet's knot fulfils the condition $\alpha_{\text{R}}=\left(\gamma-1\right)/2>\alpha$ then $\omega\,_{\text{cut, j}}$ corresponds to the upper bound of the power-law photon spectrum (the considered power-law part); otherwise, $\omega\,_{\text{cut, j}}$ is the lower bound. We assume an isotropic distribution of electrons in the knot. The index ``j'' denotes the frequency that a photon with a frequency $\omega$ has in the frame of reference of the kpc-scale jet:
\begin{equation}
\omega_\text{j}=\omega\left(1+z\right)\delta_\text{j}/\delta.
\label{eq:wj}
\end{equation}

In the second case, there is a limitation imposed by the electron spectrum, which consists in the fact that
practically all scattered radiation at a given frequency is formed due to the scattering of photons with a frequency far from the boundary values by electrons with an energy corresponding to one of the boundary values
of the power-law distribution (Fig.~\ref{fig:restrinctions}c).
Then, the spectral flux density of the scattered radiation is
\begin{equation}
F\left(\omega_\text{X}\right)=A \frac{\left[2 (1-\cos\theta_{\text{kpc}}) \right]^{\alpha+1}}{(\alpha+1) |2 \alpha-\gamma+1|}
\Gamma_{\text{cut}}^{2 \alpha-\gamma+1}\omega_{\text{X}}^{-\alpha},
\label{eq:FwxElec}
\end{equation}
where $\Gamma_{\text{cut}}$ corresponds to the lower bound of the power-law electron energy spectrum if $\alpha_{\text{R}}>\alpha$; otherwise, $\Gamma_{\text{cut}}$ is the upper bound.

The choice between formulae (\ref{eq:FwxPH}) and (\ref{eq:FwxElec}) is carried out by comparing the spectral indices of the radiation of a certain knot of the kpc-scale jet in the radio and X-ray ranges and the spectral indices of the marked power-law parts of the spectrum of the corresponding central source.
If the spectral indices of radio and X-ray radiation of the knot are close to each other,  $\alpha_{\text{R}}\approx\alpha_{\text{X}}$ hen we used expression~(\ref{eq:FwxPH}).
If $\alpha_{\text{X}}\neq\alpha_{\text{R}}$, then $\alpha_{\text{X}}$ was compared with the spectral indices of the power-law parts of the CS's spectrum, and, if $\alpha_{\text{X}}\approx\alpha_i$ expression~(\ref{eq:FwxElec}) was used with the substitution of parameters $\alpha$ and $Q$ corresponding to the $i$th part of the CS's spectrum.
For the considered sources, either the first or the second possibility was realized. The situation with $\alpha_{\text{R}}\neq\alpha_{\text{X}}$ and $\alpha_{\text{X}}\neq\alpha_i$ can be explained within IC/CS by the presence of a break in the X-ray spectrum of a knot, caused by the transition from the limitation from the photon spectrum to the limitation from the electron spectrum, as, e.g., it occurs in the high-energy spectrum of the kpc-scale jet knots near the core of the quasar PKS~1127$-$145~\citep{BP19}.

\section{THE ANGLE BETWEEN THE LINE OF SIGHT AND THE VELOCITY OF KILOPARSEC JETS}

With increasing distance from CS, the density of IC scattering photons decreases, which, with a similar density of ultrarelativistic electrons in the knots, results in a decrease in their X-ray intensity. In the knots located further than a certain distance from CS, the density of CS photons decreases to such an extent that the main source of photons for IC becomes the cosmic microwave background.
As a result of the constant density of CMB photons, the X-ray intensity of these knots is approximately the same, as in the case of the kpc-scale jet of the quasar 3C~273 \citep{Marsh01, Samb01}.
Namely, in the two knots nearest to CS, the X-ray intensity decreases with distance from CS, while in the ones it has a small, approximately constant value.
The jets of the other selected objects,excepting PKS~0637$-$752 \citep{Schwartz2000}, exhibit only a decrease in the X-ray intensity of the knots with distance from CS \citep{HoganList11, Samb04, CheungStawSiem06}.
Therefore, we assume that the X-ray radiation of all  knots of the jets of PKS~0637$-$752, PKS~1045$-$188, and PKS~1510$-$089 is formed due to IC/CS. In the farthest knots of these jets detected in the radio range, X-rays are not detected, and it is possible that IC/CMB occurs in them.

From a comparison of the fluxes of scattered radiation generated due to IC/CS with limitation from the photon spectrum and IC/CMB, it becomes possible to determine the angle between the kpc-scale jet and the line of sight \citep[(see details in][]{MBK10, BP19} or at least its lower bound \citep{BP19}:
\begin{equation}
\theta_{\text{kpc}}^{\gamma+3}\geq \frac{2^{\gamma+1}\left|2\alpha+1-\gamma\right|}{\gamma+3} W_{\text{CMB}}\frac{4 \pi c R^2}{L_{\text{CS}}} 
\left(\frac{\omega_{\text{CMB}}}{\omega_{\text{cut,\,j}}} \right)^{\left(\gamma-1\right)/2-1},
\label{eq:thetaminPH}
\end{equation}
where $\gamma$ is the spectral index of the electron energy distribution in the farthest knot, located at a projection distance $R$ from CS, the X-ray radiation of which is formed due to the IC/CS, $\omega_{\text{CMB}}$ and ${W_\text{CMB}}$ are the frequency of the maximum and the energy density of
CMB at the redshift of the considered object, respectively, and
\begin{equation}
L_{\text{CS}}=4\pi\left(1+z\right)^{3+\alpha}
\left(\frac{\delta_{\text{j}}}{\delta}\right)^{3+\alpha}  
D_L^2 Q \, \omega_{\text{cut}}^{-\alpha+1}
\label{eq:LCS}
\end{equation}
is the luminosity of CS in the reference frame of the kpc-scale jet, $D_L$ is the luminosity distance of the object, $Q$ and $\alpha$ are the coefficient of proportionality and the spectral index of the power-law part of the CS's spectrum, the scattering of photons of which makes the main contribution to the observed X-ray radiation from a given knot. Changing the sign of inequality in expression~(\ref{eq:thetaminPH}), which corresponds to the dominance of IC/CMB over IC/CS, and taking $R$ equal to the distance of the nearest knot, the X-ray radiation of which is presumably formed due to IC/CMB, we determine the upper bound of $\theta_{\text{kpc}}$. The obtained by this way angles of the kpc-scale jets of the sampled quasars are presented in Table\ref{tab:angle}.
It should be noted that, X-ray emission of the kpc-scale jets of PKS~0637$-$752, PKS~1045$-$188, and PKS~1510$-$089, using for estimation of the lower bound of the angle with the line of sight, is received from the straight part of the jets.
With increasing distance from the central source, there is a bend appears on the radio maps behind the straight part of the jet and the angle with the line of sight of the jet part after bend can be greater.

\begin{table*}
 \caption{
Angle with the line of sight and velocity of kiloparsec-scale
jets of quasars.
 } 
 \label{tab:angle} 
 \medskip
 \begin{tabular}
 {|c|c|c|c|c|}
  \hline
 Object & $\theta_{\rm kpc}$, deg & $\delta_{\rm kpc}$ & $\beta_{\rm kpc}$ & $\Gamma_{\rm kpc}$ \\
  \hline
 \multirow{4}{*}{PKS~0637$-$752} &  \multirow{2}{*}{27\footnote{The lower bound of the angle $\theta_{\rm kpc}$.}} & \multirow{2}{*}{2.19} & 0.87 & 2.0 \\
            &   &  & 0.91 & 2.4 \\
\cline{2-5}
            &  \multirow{2}{*}{37\footnote{The value of $\theta_{\rm kpc}$, exceeding by $10^\circ$ the minimum; it is given for comparison.}} & \multirow{2}{*}{1.65} & 0.56 & 1.2 \\
            &   &  & 0.98 & 4.7 \\
\hline
\multirow{4}{*}{PKS~1045$-$188} &  \multirow{2}{*}{34$^\text{\textit{a}}$} & \multirow{2}{*}{1.79} & 0.81 & 1.7 \\
            &   &  & 0.84 & 1.9 \\
\cline{2-5}
            &  \multirow{2}{*}{44$^\text{\textit{b}}$} & \multirow{2}{*}{1.44} & 0.70 & 1.4 \\
            &   &  & 0.74 & 1.5 \\
\hline          
 \multirow{2}{*}{3C 273} &  \multirow{2}{*}{25$-$26} & \multirow{2}{*}{2.37} & 0.90 & 2.3 \\
            &   &  & 0.91 & 2.4 \\
\hline
 \multirow{4}{*}{PKS 1510$-$089} &  \multirow{2}{*}{24$^\textit{a}$} & \multirow{2}{*}{2.43} & 0.88 & 2.1 \\
            &   &  & 0.94 & 2.9 \\
\cline{2-5}
            &  \multirow{2}{*}{34$^\textit{b}$} & \multirow{2}{*}{1.77} & 0.598 & 1.3 \\
            &   &  & 0.984 & 5.6 \\
            
\hline
PKS~1127$-$145\footnote{The angle
with the line of sight and the velocity of the kpc-scale jet were
determined by \citet{BP19} and are presented here for comparison.} & 35  & 1.74  & 0.8  & 1.7  \\
\hline
  \end{tabular}
\end{table*}

It is worth noting that the angles $\theta_{\text{kpc}}$ are several dozen degrees, while the angles with the line of sight of the pc-scale jets of these sources are not greater than $7^\circ$ (see Table~\ref{tab:info}).
This difference can be explained by the effect of the relativistic aberration, when a truly straight jet slows down between the pc- and kpc-scales \citep{BP19}.
Then, the Doppler factor of the kpc-scale jets, $\delta_{\text{kpc}}$, can be estimated, the value of which on average for all sources is 2 (see Table~\ref{tab:angle}).
Knowing $\delta_{\text{kpc}}$ and $\theta_{\text{kpc}}$, one can estimate the velocity of the kpc-scale jets, $\beta_{\text{kpc}}$, the values of which mainly lie in the range from $\approx0.7$ to 0.9, and the corresponding Lorentz factor is $\Gamma_{\text{kpc}}\approx1.2-2.9$ (see Table~\ref{tab:angle}).

\section{PHYSICAL PARAMETERS OF KILOPARSEC JETS}

The relativistic electrons belonging to the same power-law energy distribution generate synchrotron radiation in the radio range and, through IC, in the X-ray range. 
Therefore, by comparing the spectral fluxes at the radio and X-ray frequencies, it is possible to estimate the perpendicular to the line of sight component of the magnetic field \citep{BP19}:
\begin{equation}
B_\bot=\biggl[ \frac{32 \pi^2 D_L^2}{c \sigma_T V \mathcal{K}} \left(1+z \right)^{3+\alpha_\text{R}}\left(m_e c^2 \right)^{\gamma-1}  
\left(0.29\frac{3 e}{2 m_e c} \right)^{1-\alpha_\text{R}} \omega_\text{R}^{\alpha_\text{R}} F_{R}\biggr]^{1/\left(1+\alpha_\text{R} \right)},
\label{eq:Bperp}
\end{equation}
where $\sigma_T$ is the Thomson scattering cross section, $e$ is the electron charge, $\alpha_{\text{R}}$ is the spectral index of radio emission, and $F_{\text{R}}$ is the observed flux at the frequency $\omega_{\text{R}}$.
The normalization factor $\mathcal{K}$ of the electron spectrum is expressed from the observed spectral flux in the X-ray range by formulae~(\ref{eq:FwxPH}) or (\ref{eq:FwxElec}), that makes the estimate of $B_\bot$ independent of the emission region volume and of the density of emitting electrons.
The values of $B_\bot$ determined in this way turn out to be one or two orders of magnitude smaller than the values of the magnetic field corresponding to the equipartition condition of energy between the emitting electrons and the magnetic field (see Tables~\ref{tab:par0637}-\ref{tab:par1045}).
In this case, the values of are approximately two times greater than the magnetic field determined from IC/CMB.

In the case of IC/CS with a limitation imposed by the photon spectrum, the formula for the spectral flux of scattered radiation (\ref{eq:FwxPH}) includes one of the boundaries of the power-law energy spectrum of scattering photons. This frequency can be determined from the observed CS's spectrum (see formula (\ref{eq:wj})).
While, the expression for $F\left(\omega_{\text{X}} \right)$ under the limitation imposed by electron spectrum (\ref{eq:FwxElec}) includes one of its boundaries, which is unknown. Some estimates for the boundaries of the electron spectrum can be obtained from synchrotron radio, infrared, and optical radiation and from IC/CS using the relation between the energies of interacting particles,
\begin{equation}
\Gamma^2 = \frac{\omega_{\text{X}}}{k_{\text{IC}}\, \omega_{\text{j}}},
\label{eq:osv}
\end{equation}
where $k_{\text{IC}}=4/[3(1+z)]$. Let us consider each object in
more detail.

\subsection{PKS~0637---752}

\begin{figure}
\center{ \includegraphics[scale=0.5]{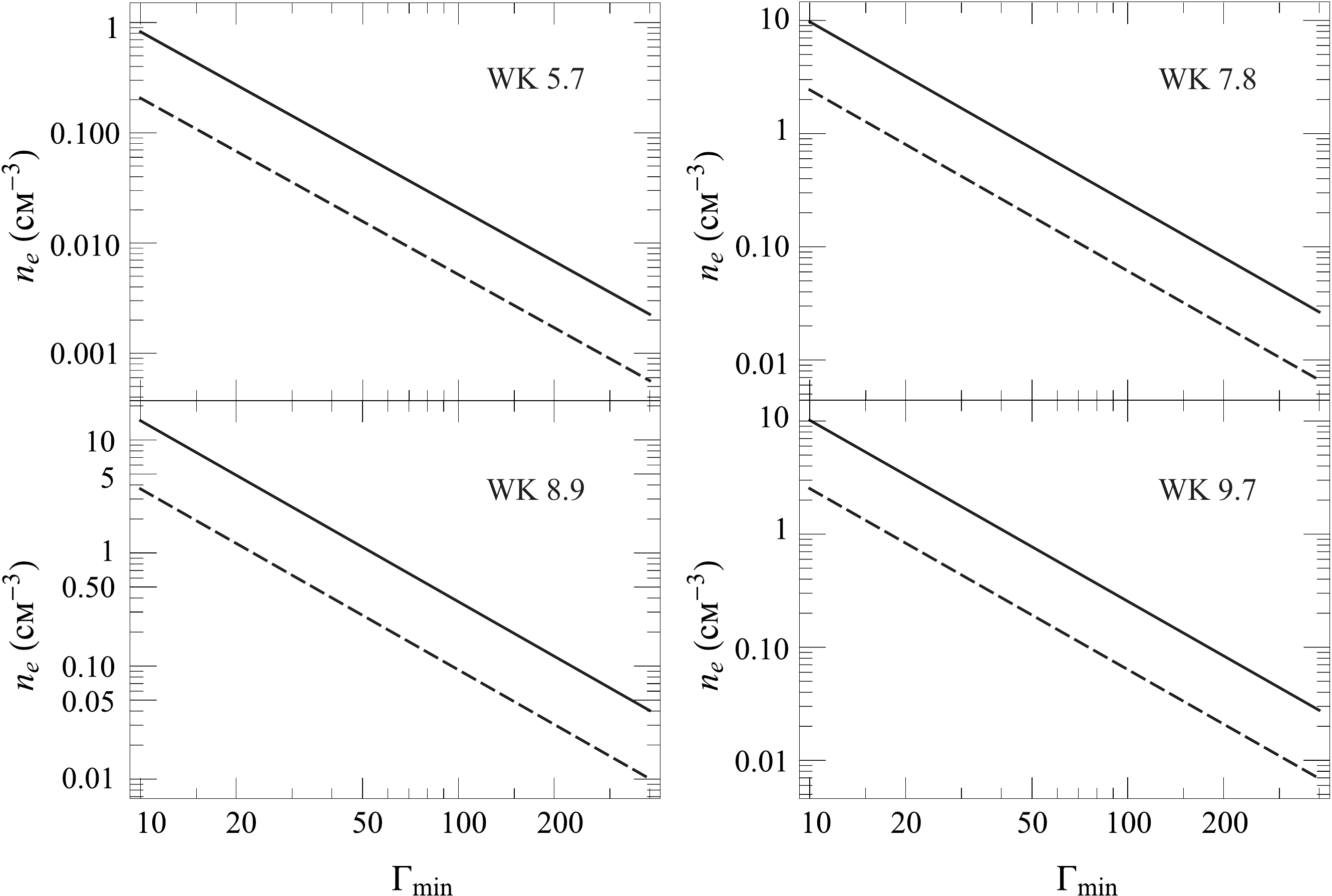}  }  
 \caption{
 Number density of emitting electrons in the knots of the PKS~0637$-$752 jet vs. $\Gamma_{\text{min}}$.
  Solid lines represent the lower limit for the angle between the
kpc-scale jet and the line of sight $\theta_{\text{kpc}}=24^\circ$. Dashed lines correspond to case of $\theta_{\text{kpc}}=34^\circ$.  }
 \label{fig:0637ne}
\end{figure}

\begin{figure}
\center{ \includegraphics[scale=0.7]{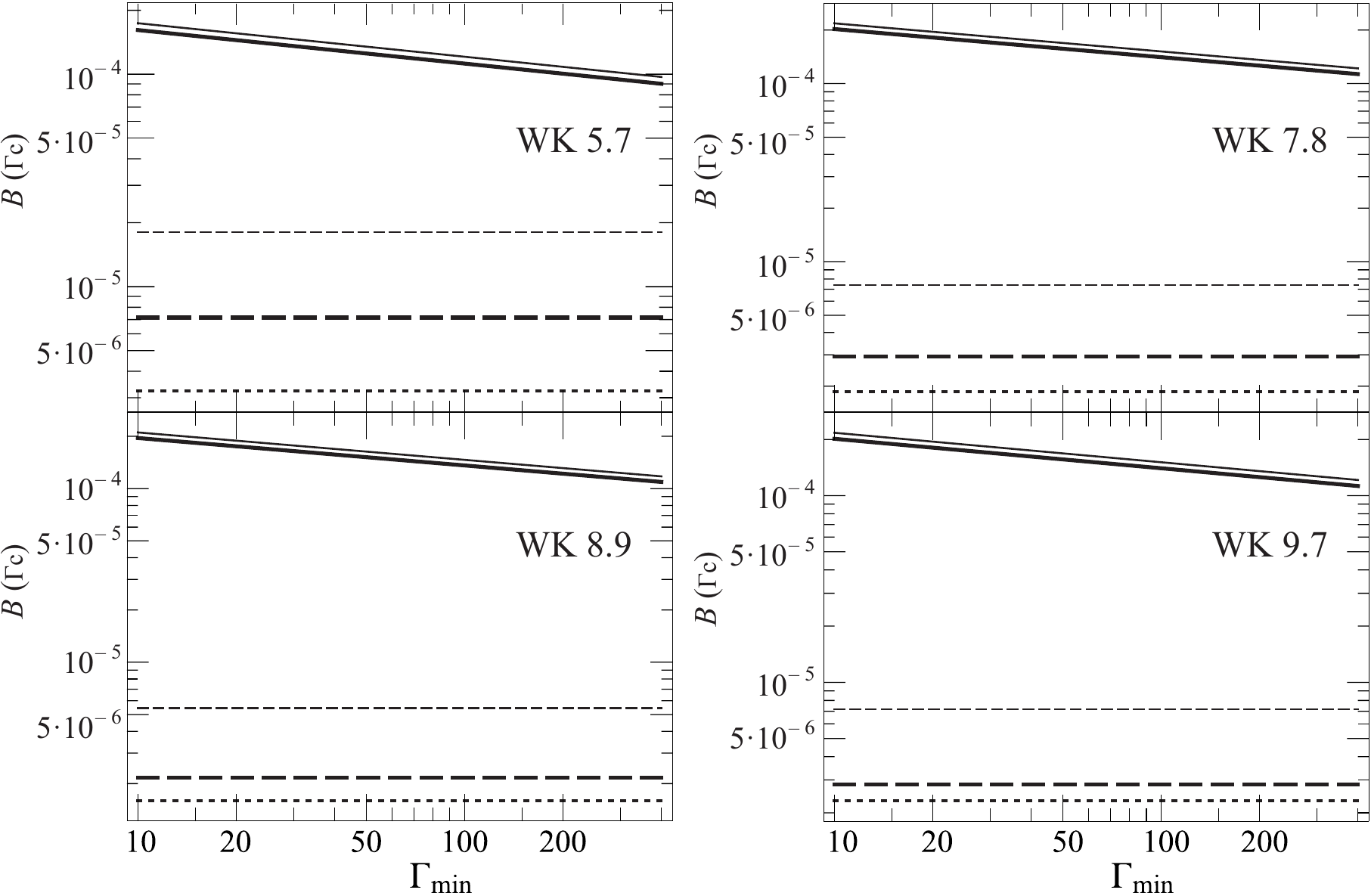}  }  
 \caption{
 Magnetic field in the knots of the PKS~0637$-$752 jet.
 Thick and thin lines correspond to $\theta_{\text{kpc}}=27^\circ$ and $\theta_{\text{kpc}}=37^\circ$, respectively. Solid lines represent to the magnetic field strength under fulfilment of the energy equipartition condition. Dashed lines mark the magnetic field strengths obtained from the comparison of radio and X-ray fluxes in the framework of IC/CS. Dotted lines denote the magnetic field determined under the assumption of IC/CMB. All values of the magnetic field are shown for $\delta_{\text{kpc}}=1$.
 }
 \label{fig:0637H}
\end{figure}

The morphology of the kpc-scale jet of the quasar PKS~0637$-$752 is thoroughly described by \citet{Chartas00, Schwartz2000}.
In the radio range, the jet is directed to the west and, starting from the distance $\approx10^{\prime\prime}$ from the core, bends to the north. To the east of the core there is a counter-lobe.
X-rays are detected from the jet knots WK~5.7, WK~7.8, WK~8.9, and WK~9.7, located at a distance of $5-10^{\prime\prime}$ from the core\footnote{The nomenclature reflects the belonging of the knots to the jet
located west of the core and their angular distance from the core.}.
In the optical range, knots WK~7.8, WK~8.9, WK~9.7 are found \citep{Mehta09}; for the first two of them, infrared radiation was detected \citep{Uchiyama05}.
For the region from $3^{\prime\prime}$ to $10^{\prime\prime}$ from the core, the spectral index in the radio
range is $\alpha_{\text{R}}=0.81$ \citep{Chartas00}.
In the X-ray range, for the region located within $4.0-6.^{\prime\prime}5$ from the core, the spectral index is $\alpha_{\text{X}}=1.1\pm0.3$, while, for the region centered at the WK~8.9 knot and having a radius of $2.^{\prime\prime}5$, $\alpha_{\text{X}}=0.85\pm0.1$ \citep{Chartas00}.
Since, for the knots WK~7.8, WK~8.9, and WK~9.7, within the errors, the spectral indices of radio and X-ray radiation are equal, $\alpha_{\text{R}}=\alpha_{\text{X}}$, then in these knots, the X-ray radiation is formed due to IC of photons belonging to the second part of the CS spectrum under the limitation from the photon spectrum.
Since, for the spectral index of the radiation of the second part of the CS spectrum is  $\alpha_2>\alpha_\text{R}$, expression ~(\ref{eq:FwxPH}) implies that the main contribution to the X-ray radiation at the observed frequency $\omega_{\text{X}}=1.5\times10^{18}$~s$^{-1}$ (photon energy 1~keV) is made by IC of photons with a frequency corresponding to the lower bound (see Table~\ref{tab:CSspectra}).
In this case, scattering occurs on electrons belonging to a power-law distribution with a spectral index $\gamma=2\alpha_{\text{R}}+1\approx2.6$.
We denote by $\omega_{0, {\text{j}}}$ the frequency corresponding to the upper boundary of the 1st part and the lower boundary of the 2nd part of the CS spectrum in the reference frame of the kpc-scale jet. Then, as can be seen from Fig.~\ref{fig:restrinctions}b, the lower boundary of the electron spectrum, $\Gamma_{\text{min}}$, is smaller than the Lorentz factor of electrons, scattering photons with $\omega_{0, {\text{j}}}$ to the frequency $\omega_{\text{X}}$.
Expression~(\ref{eq:osv}) implies that $\Gamma_{\text{min}}<400$.
We took this into account in further estimates of the magnetic field strength and the density of emitting electrons. These estimates were carried out using observational data \citep{Chartas00, Mehta09} and formulae~(\ref{eq:FwxPH}) and (\ref{eq:Bperp}) for the upper bound of
the knots size of $0.^{\prime\prime}4$ \citep{Chartas00}. 
The magnetic fields under the energy equipartition condition and under the assumption of IC/CMB were estimated using expressions~(9) in \citep{BP19} and (10) in \citep{MBK10} (corrected by $z$ and a
factor of 0.29), respectively.
The obtained values of these parameters are presented in Table~\ref{tab:par0637} and illustrated by Fig.~\ref{fig:0637ne} and Fig.~\ref{fig:0637H}. 
If the actual size of the emission region is smaller than $0.^{\prime\prime}4$, then the values of the magnetic field and electron density are somewhat greater.

\begin{table*}
 \caption{
Physical parameters for the knots of the the PKS~0637$-$752 quasar jet.
The columns represent: (1) knot; (2) $\theta_{\rm kpc}$ used in calculations; (3, 4) electron densities for $\Gamma_{\rm min}=10$ and $\Gamma_{\rm min}=100$; (5) the magnetic field determined from the flux ratio of radio to X-ray; (6) the magnetic field determined from the radio emission of the knot, necessary to fulfil the energy equipartition condition for $\Gamma_{\rm min}=10$; (7)  the magnetic field determined under the assumption of IC/CMB for $\delta_{\rm kpc}=1$; (8) the boundaries of the power-law electron distribution.
 } 
 \label{tab:par0637} 
 \medskip
 \begin{tabular}
 {|c|c|c|c|c|c|c|l|}
  \hline
Knot & $\theta_{\rm kpc}$, deg & $n_{e10}$, cm$^{-3}$ & $n_{e100}$, cm$^{-3}$ & $B_{\perp}$, G & $B_\text{eq}$, G & $B_{\perp, \rm CMB}$, G & $e^{-}$ spectrum\\
  \hline
(1) & (2) & (3) & (4) & (5) & (6) & (7) & (8) \\
\hline
 \multirow{2}{*}{WK5.7} &  27 & 0.83 & 0.02 & $7.2 \cdot 10^{-6}$ & $1.6 \cdot 10^{-4}$ & \multirow{2}{*}{$3.2 \cdot 10^{-6}$} & \\
 \cline{2-6}
            & 37  & 0.21  & 0.01  & $1.8 \cdot 10^{-5}$  & $1.7 \cdot 10^{-4}$  & &  \\
\cline{1-7}
 \multirow{2}{*}{WK7.8} &  27 & 9.72 & 0.24 & $2.9 \cdot 10^{-6}$ & $2.0 \cdot 10^{-4}$ & \multirow{2}{*}{$1.9 \cdot 10^{-6}$} & \multirow{2}{*}{ \raisebox{-2.5ex}[0cm][0cm]{$\Gamma_{\text{br}}<5\cdot10^5$}}  \\
 \cline{2-6}
            & 37  & 2.43  & 0.06  & $7.4 \cdot 10^{-6}$  & $2.2 \cdot 10^{-4}$  & & \\
\cline{1-7}
 \multirow{2}{*}{WK8.9} &  27 & 14.00 & 0.37 & $2.2 \cdot 10^{-6}$ & $2.0 \cdot 10^{-4}$ & \multirow{2}{*}{$1.6 \cdot 10^{-6}$} & \multirow{2}{*}{ \raisebox{2.0ex}[0cm][0cm]{$\Gamma_{\text{min}}<400$}} \\
 \cline{2-6}
            & 37  & 3.69  & 0.09  & $5.5 \cdot 10^{-6}$  & $2.1 \cdot 10^{-6}$  & & \\
\cline{1-7}
 \multirow{2}{*}{WK9.7} &  27 & 10.15 & 0.26 & $2.8 \cdot 10^{-6}$ & $2.0 \cdot 10^{-4}$ & \multirow{2}{*}{$2.3 \cdot 10^{-6}$} & \\
 \cline{2-6}
            & 37  & 2.53  & 0.06  & $7.2 \cdot 10^{-6}$  & $2.2 \cdot 10^{-4}$  & & \\
\hline
  \end{tabular}
\end{table*}

For the knot WK~5.7, within the measurement errors, we have $\alpha_{\text{X}}=\alpha_{\text{R}}$.
Therefore, X-ray radiation is formed due to IC of the 2nd part of the CS spectrum by electrons with an energy corresponding to one of the boundaries of the power-law distribution (see formula~(\ref{eq:FwxElec})).
Since  $\alpha_{\text{R}}<\alpha_2$, this is the upper boundary.
If the X-ray spectrum measured in the frequency range from $\omega_1$ to $\omega_2$ (corresponding to the photon energies from 0.2 to 4~keV) is power-law, then electrons with the Lorentz factor $\Gamma_{\text{max}}$ must scatter photons with a frequency $\omega_{0, {\text{j}}}$ to frequencies lower than $\omega_1$.
Using~(\ref{eq:osv}), we obtain $\omega_1>k_{\text{IC}}\,\omega_{\text{0, j}}\Gamma^2_{\text{max}}$, whereas the photons with a frequency  $\omega_{\text{max, j}}$, corresponding to the upper boundary of the 2nd part of the spectrum, are scattered to frequencies greater than  $\omega_2<k_{\text{IC}}\,\omega_{\text{max, j}}\Gamma^2_{\text{max}}$. 
Thus, the range of possible values is $\Gamma^2_{\text{max}}=36-189$, which is very small, since electrons in a magnetic field, e.g., of $\sim10^{-5}$~G \citep{Harris06} cannot create the observed synchrotron
radiation.

Then, in the indicated range of X-ray radiation, the spectrum, possibly, has a break, associated with the transition from the limitation from the photon spectrum to the limitation from the electron spectrum with an increase in the observation frequency. This break will occur at a frequency corresponding to the scattering of photons with  $\omega_{0, {\text{j}}}$ by electrons with $\Gamma_{\text{max}}$.
Using~(\ref{eq:osv}), we found that $\sqrt{\omega_1/(k_{\text{IC}}\,\omega_{\text{0, j}})}=189<\Gamma_{\text{max}}<847=\sqrt{\omega_2/(k_{\text{IC}}\,\omega_{\text{0, j}})}$.
From the knot WK~5.7, radio emission at a frequency of 8.6~GHz was detected \citep{Chartas00}.
For synchrotron radiation at this frequency by electrons with $\Gamma_{\text{max}}=840$, the required magnetic field strength is 0.01~G. This magnitude of the magnetic field corresponds to parsec-scale jets \citep[(see, e.g.,][]{Pushkarev12}, but is very large for kiloparsec-scale jets \citep{Harris06}.
Possibly, the electron energy spectrum in the knot WK~5.7 has a break similar to the break in other knots. The smaller value of the Lorentz factor at which the break occurs in comparison with other knots can be interpreted by the fact that the knot WK~5.7 is closer to CS and, therefore, the IC/CS losses are greater. 
Then, the spectrum in the radio range must be steeper. Based on the available data on the radio spectrum of the entire western jet up to its bend, it is difficult to draw any conclusions about the radio spectrum of the knot WK~5.7.

On the other hand, $\alpha_{\text{R}}=\alpha_{\text{X}}$ within the measurement errors. Then, in the knot WK~5.7, as well as in the other ones, IC occurs in the 2nd part of the CS spectrum with limitation from the photon spectrum.
The electron density and the magnetic field strength found in this case are given in Table~\ref{tab:par0637}.

Let us consider the spectrum of scattered radiation at frequencies exceeding the \textit{Chandra} operating range. With increasing $\omega_{\text{X}}$, the spectrum of scattered radiation $\alpha_{\text{X}}=\left( \gamma-1 \right)/2$ will persist up to the frequency
\begin{equation}
\omega_{\text{X, br}}=k_{\text{IC}}\, \omega_{\text{0, j}} \Gamma^2_{\text{br}},
\label{eq:wxbr}
\end{equation}
where $\Gamma_{\text{br}}$ is the upper boundary of the electron energy distribution with the index $\gamma=2.6$. Synchrotron
infrared and optical radiation detected from the knots WK~7.8, WK~8.9, and WK~9.7 is characterized by the spectral index  $\alpha_{\text{Opt}}^{\text{IR}}\approx1.6$~\citep{Uchiyama05}, which indicates a break in the electron energy spectrum, most likely arising from the radiation losses of the electron energy, which are most significant for electrons with higher energy.
Electrons with a Lorentz factor smaller than the Lorentz factor $\Gamma_{\text{br}}$, at which the break occurs, emit due to the synchrotron mechanism in the radio range, and electrons with a Lorentz factor greater than $\Gamma_{\text{br}}$ emit in the infrared and optical ranges. Based on this, one can estimate $\Gamma_{\text{br}}$. For example, with a magnetic field of $5\cdot10^{-6}$~G the Lorentz factor of the break is within $2\cdot 10^4-7\cdot10^5$. 
The IC/CS by electrons with $\Gamma>\Gamma_{\text{br}}$ does not make a significant contribution to the scattered radiation up to the frequency $\omega_{\text{X, br}}$; otherwise, it would be reflected in the value of $\alpha_{\text{X}}$.

\begin{figure}
\center{ \includegraphics[scale=0.7]{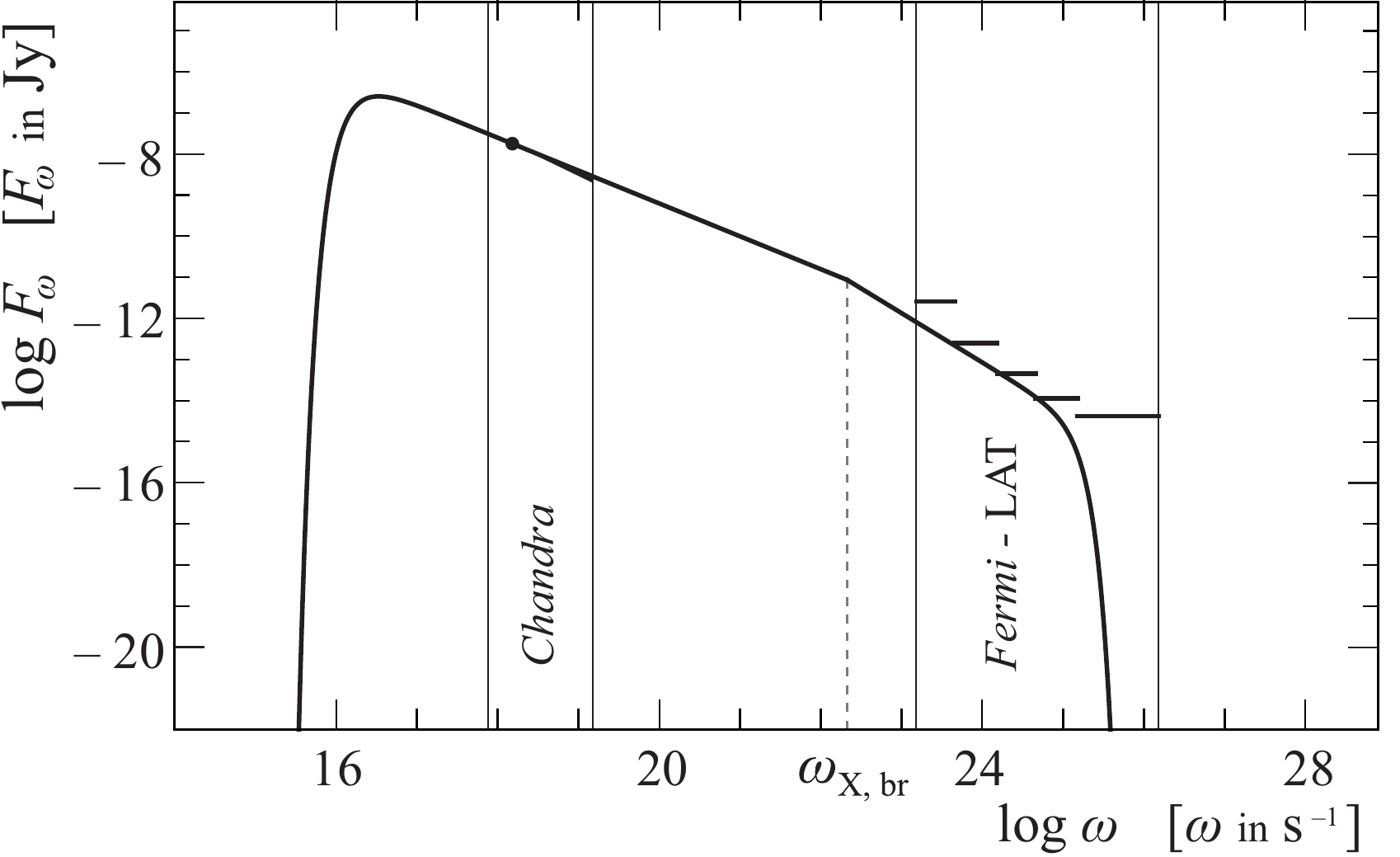}  }  
 \caption{
 High-frequency spectrum of total radiation of quasar PKS~0637$-$752 kpc-scale jet knots, which were detected in the X-ray range, simulated in the framework of IC/CS: (dot) total observed X-ray flux; (filled triangle) the measurements errors of the spectral
index, (horizontal segments) the upper limits of the total flux of the kpc-scale jet \citep{Meyer15}, (dotted line) the frequency of the spectrum break, corresponding to the transition from the limitation imposed by the photon spectrum to the limitation imposed
by the electron spectrum at $\Gamma_{\rm br}=5\cdot10^{4}$; (solid vertical lines) the \textit{Chandra} and \textit{Fermi}-LAT operating ranges.
 }
 \label{fig:0637gamma}
\end{figure}

At frequencies higher than $\omega_{\text{X, br}}$, the total flux of scattered radiation with taking into account the relation specified by expression~(\ref{eq:osv}) has two components: namely, IC of the 2nd part of the CS spectrum by electrons with  $\Gamma_{\text{br}}$ and IC of photons with a frequency $\omega_{\text{0,j}}$ by electrons with $\Gamma>\Gamma_{\text{br}}$. 
Substituting into expressions~(\ref{eq:FwxElec}) and (\ref{eq:FwxPH}) the corresponding quantities and taking into account the continuity in the total distribution of electrons, one can make sure that the contribution of IC by the high-energy electron distribution to the scattered radiation is insignificant. The total spectrum of the scattered radiation generated in the knots WK~5.7, WK~7.8, WK~8.9, and WK~9.7, plotted with the parameters $\Gamma_{\text{br}}=5\cdot10^4$ and $\Gamma_{\text{min}}=50$, shown in Fig.~\ref{fig:0637gamma}.
The IC of photons with a frequency $\omega_{\text{0,j}}$ by electrons with $\Gamma_{\text{min}}$ and the IC of photons with
a frequency $\omega_{\text{max, j}}$ by electrons with $\Gamma_{\text{br}}$ determine, according to expression~(\ref{eq:osv}), the low- and high-energy cutoffs of the high-frequency spectrum of the knots,respectively. 
It can be seen that the assumption about IC/CS as a mechanism of high-frequency radiation of the kpc-scale jet of the quasar PKS~0637$-$725 cannot be refuted by the available observational data in the gamma-ray range \citep{Meyer15}.

\subsection{3C~273}

\begin{figure}
\center{ \includegraphics[scale=0.7]{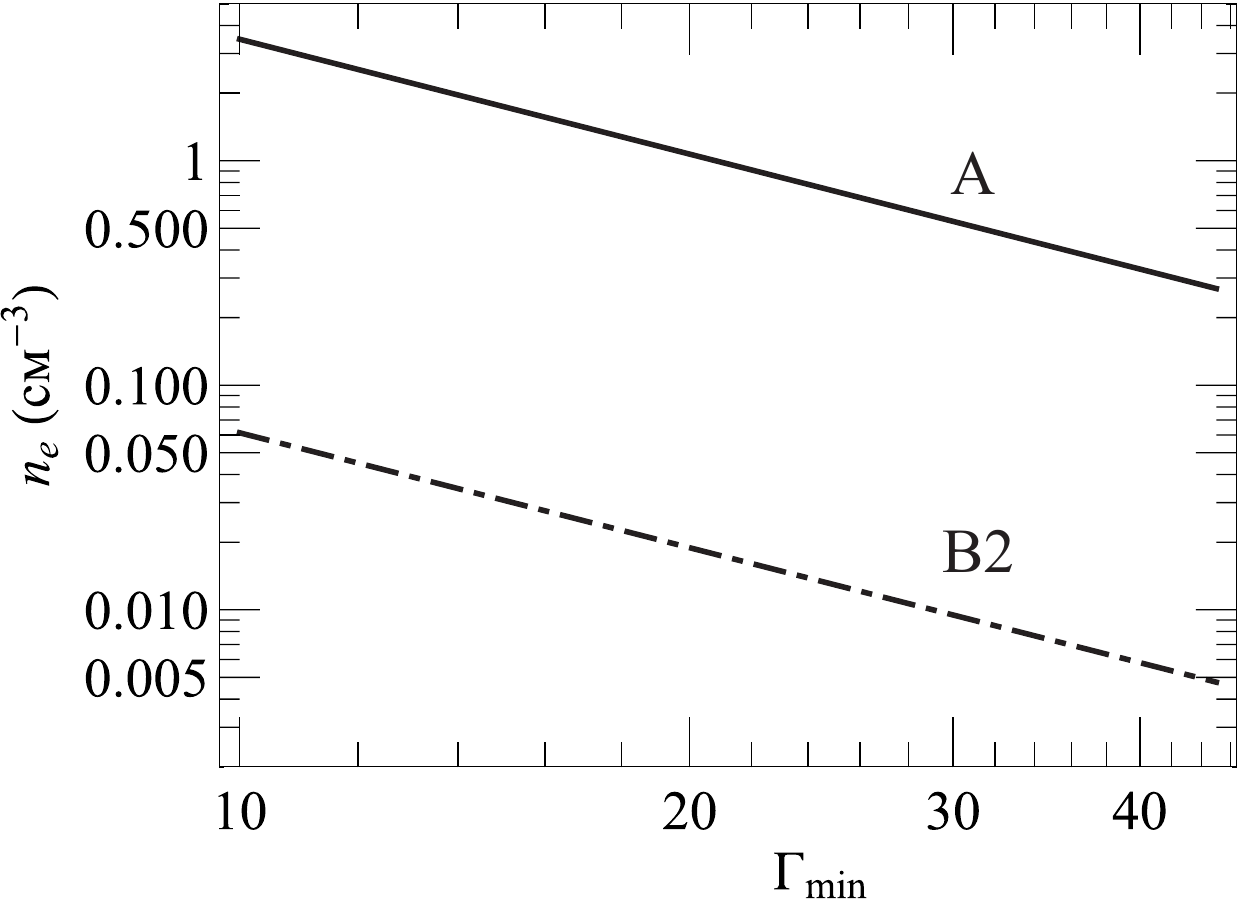}  }  
 \caption{
The number density of emitting electrons in the nearest to CS knots~A and B2 of the 3C~273 quasar jet. The values were obtained within IC/CS and $\theta_{\rm kpc}=27^\circ$.
 }
 \label{fig:3c273ne}
\end{figure}

\begin{figure}
\center{ \includegraphics[scale=0.7]{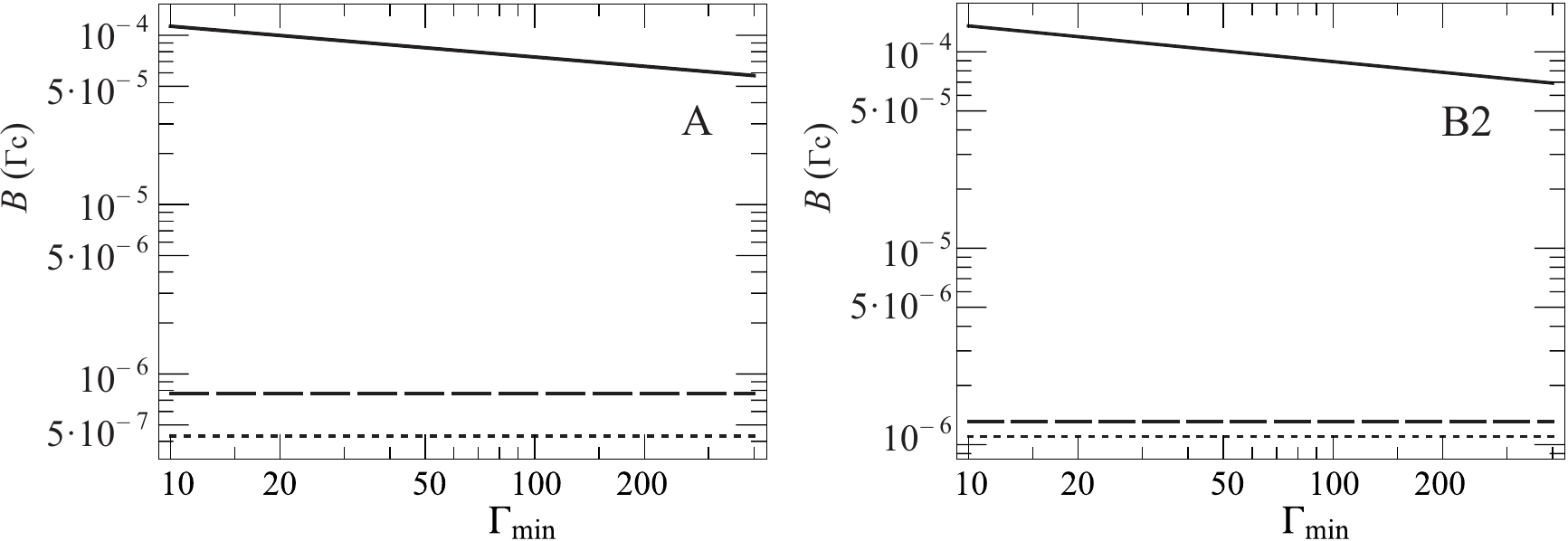}  }  
 \caption{
Magnetic field in the nearest to CS knots of the 3C~273 jet: (solid line) the magnetic field strength under the energy equipartition
condition; (dashed lines) the magnetic field strength obtained from a comparison of radio and X-ray fluxes within the
IC/CS; (dotted line) the magnetic field determined under the assumption of IC/CMB. All values of the magnetic field are shown
for $\delta_{\text{kpc}}=1$ and $\theta_{\rm kpc}=27^\circ$.
 }
 \label{fig:3c273H}
\end{figure}

One of the nearest quasars with an extended onesided kpc-scale jet (up to 23$^{\prime\prime}$), 3C~273 was intensely observed in radio \citep[see, e.g.,][]{}(ConwayGP93, PerleyMeis17), infrared \citep{Uchiyama06}, optical \citep{Bahcall95,Jester01}, ultraviolet \citep{Jester05}, and X-ray \citep{HarSt87, Roser00, Marsh01, Samb01} ranges. 
In the radio range, the jet knots are found up to $\approx23^{\prime\prime}$ from the core.
Starting from the distance of $\approx12^{\prime\prime}$ from the core, where the jet begins to be detected at other frequencies, the radio intensity sharply increases and continues to increase with distance, reaching a maximum in the hot spot located at $22^{\prime\prime}$. 
In the optical range, the maximum intensity is also observed in the region corresponding to the hot spot, but approximately $1^{\prime\prime}$ closer to the core than the peak intensity in the radio range.
For the rest knots, the optical intensity is approximately the same \citep[see, e.g.,][]{Jester01}. 
In the X-ray range, the intensity is maximal in the knot~A\footnote{Here, we adhere to the nomenclature of knots used by \citep{Jester01}.}, nearest to the core; 
then it decreases by half at the knot~B2\footnote{Note that we do not distinguish the region located between the knots~A and B2 as a separate knot, since it has a low intensity at all observed frequencies and the position of its peak brightness strongly depends on frequency. This behavior requires additional explanation, which is beyond the scope of this work.}; 
and, in subsequent knots, it decreases by almost an order of magnitude and has an approximately constant value. 
Farther than $21^{\prime\prime}$ from the core, X-ray radiation is not detected \citep[see][]{Marsh01, Samb01,Jester06}.
\citet{MBK10} assumed that, for the knots~A and B2, X-ray radiation is generated due to IC/CS and, in other knots, via IC/CMB. 
The spectral indices of the knots~A and B2 in the radio and X-ray ranges are $\approx0.85$ \citep[according to the data of][]{ConwayGP93, Jester07} and differ from the spectral indices of the CS radiation (see Table~\ref{tab:CSspectra}).
Therefore, according to expression~(\ref{eq:FwxPH}), the main contribution to the observed X-ray flux comes from the scattering of photons with a frequency $\omega_{\text{max,\,j}}$ by electrons
with energies far from the boundary values. 
Formula~(\ref{eq:osv}) implies that $\Gamma_{\text{min}}<50$. 
The results of the estimates of the electron number density and the magnetic field strength are presented in Table~\ref{tab:par273} and Figs.~\ref{fig:3c273ne} and~\ref{fig:3c273H}.

\begin{table*}
 \caption{
Physical parameters of the knots A and B2 of the quasar 3C 273 jet, in which IC/CS occurs.
The columns represent (1) the knot; (2) $\theta_{\rm kpc}$ used in calculations; (3), (4) electron number density for $\Gamma_{\rm min}=10$ and $\Gamma_{\rm min}=40$; (5) the magnetic field
determined from the ratio of fluxes in the radio and X-ray ranges; (6) the magnetic field determined from the radio emission of the knot,
which is necessary to fulfil the energy equipartition condition for $\Gamma_{\rm min}=10$; (7) the magnetic field determined under the assumption of IC/CMB for $\delta_{\rm kpc}=1$; (8) boundaries of the power-law electron distribution.
 } 
 \label{tab:par273} 
 \medskip
 \begin{tabular}
 {|c|c|c|c|c|c|c|l|}
  \hline
Knot & $\theta_{\rm kpc}$, deg & $n_{e10}$, cm$^{-3}$ & $n_{e100}$, cm$^{-3}$ & $B_{\perp}$, G & $B_\text{eq}$, G & $B_{\perp, \rm CMB}$, G & $e^{-}$ spectrum \\
  \hline
(1) & (2) & (3) & (4) & (5) & (6) & (7) & (8) \\
\hline
A  & \multirow{2}{*}{27} & 3.47 & 0.33 & $6.1 \cdot 10^{-7}$ & $2.5 \cdot 10^{-3}$ & $4.3 \cdot 10^{-7}$ & \raisebox{-0.5ex}[0cm][0cm]{$\Gamma_{\text{min}}<50$} \\
\cline{1-1} \cline{3-7}
B2  &   & $6.1 \cdot 10^{-2}$ & $5.8 \cdot 10^{-3}$ & $1.2 \cdot 10^{-5}$ & $3.0 \cdot 10^{-3}$ & $7.5 \cdot 10^{-6}$ & \raisebox{0.7ex}[0cm][0cm]{$\Gamma_{\text{max}}\leq 10^5$ }\\ 
\hline
  \end{tabular}
\end{table*}

\begin{figure}
\center{ \includegraphics[scale=0.6]{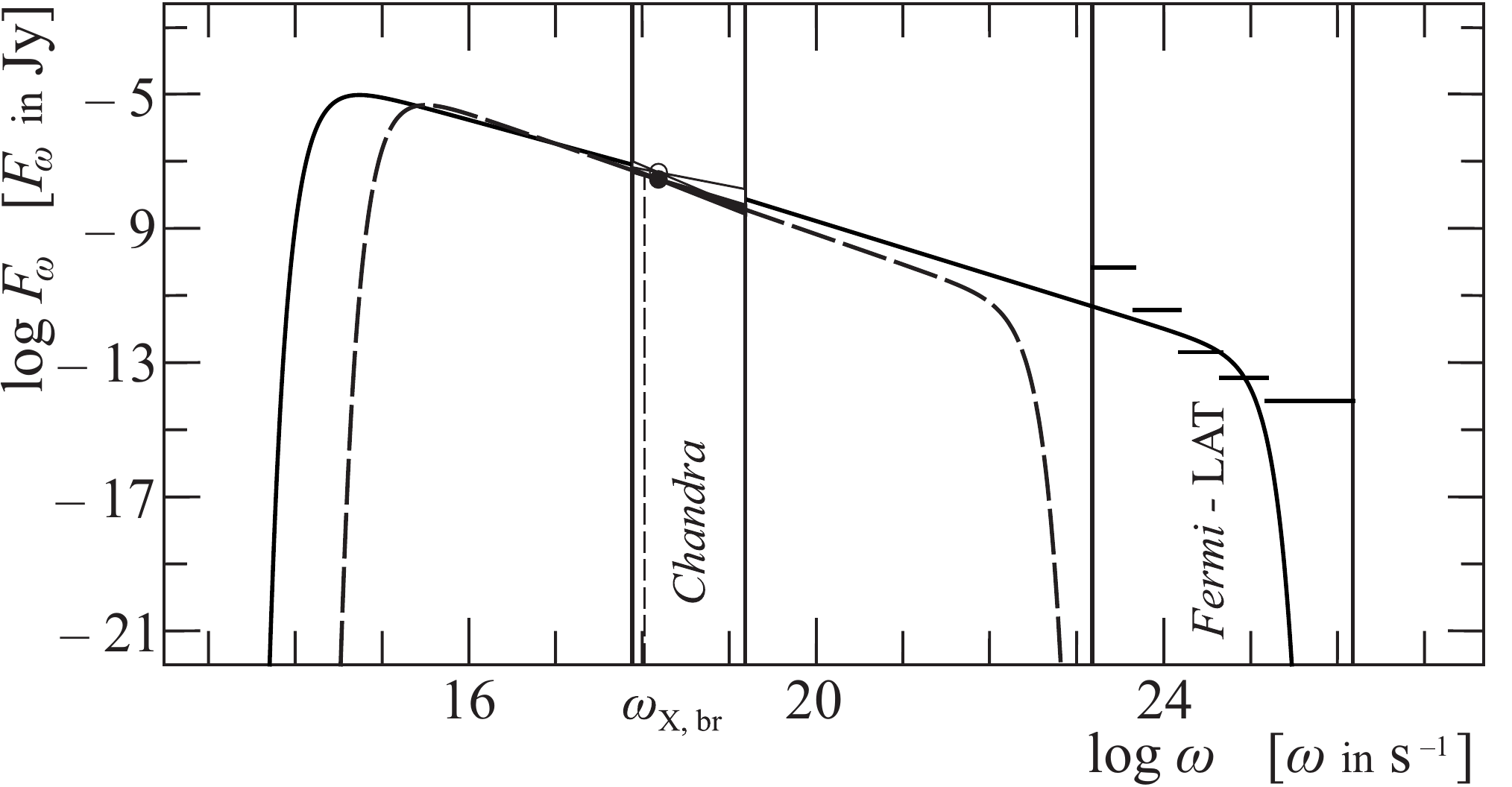}  }  
 \caption{
 High-frequency spectra of the inner (solid line) and outer (dashed line) parts of the quasar 3C~273 jet. By the inner part
of the jet we mean the knots~A and B2, the X-ray radiation of which is generated by IC/CS. The observed X-ray flux is shown in
white. The break in the spectrum at the frequency $\omega_{\text{X, br}}$ is caused by the transition from the limitation imposed by the electron
spectrum to the limitation imposed by the photon spectrum. The outer part includes the rest of the knots the X-ray radiation of
which is generated via IC/CMB. The errors of determination of the spectral index is indicated by a filled triangle. The horizontal
line segments mark the upper bounds of the fluxes from the entire kpc-scale jet \citep{MeyGeor14}. The solid vertical lines mark the \textit{Chandra} and \textit{Fermi}-LAT operating ranges.
 }
 \label{fig:3c273gamma}
\end{figure}

In the high-frequency spectrum of the knots~A and B2 (Fig.~\ref{fig:3c273gamma}), as well as in the spectrum of the knots of the PKS~0637$-$752 jet, at the frequency
\begin{equation}
    \omega_{\text{X, br}}=k_{\text{IC}}\,\omega_{\text{max, j}}\Gamma_{\text{min}}^2
\end{equation}
there is a break caused by the transition from the limitation from the electron spectrum to the limitation from the photon spectrum. The low-energy boundary of the high-frequency spectrum is determined by IC on electrons with the Lorentz factor $\Gamma_{\text{min}}$.
Only for the knots~A and B2, the frequency of scattered photons is $\omega_{\text{0, j}}=1.8\cdot10^{11}$~s$^{-1}$; for other knots this frequency is $\omega_{\text{CMB}}=1.2\cdot10^{12}$~s$^{-1}$.
Since $\omega_{\text{0, j}}<\omega_{\text{CMB}}$, the low-frequency cutoff of the spectrum of the knots~A and B2 is at a lower frequency than that for the distant knots. 
This explains the observed difference in the energy distribution in the spectrum of the near and far knots \citep{Jester07} by the fact that, in the infrared and optical radiation of the knots~A and B2, a significant contribution is made by the radiation due
to IC/CS.
For example, the IC of the 2nd part of the CS spectrum by electrons with $\Gamma_{\text{min}}$ (see expression (\ref{eq:FwxElec})) can reproduce the observed flux in the optical range for the near knots with an accuracy better than $10\%$ and give $\approx50\%$ of the flux in the infrared range for $\Gamma_{\text{min}}=40$. 
For $\Gamma_{\text{min}}=10$ IC/CS will give up to $70\%$ of the observed flux in the infrared range, whereas, for the distant knots, the X-ray spectrum generated via the IC/CMB is cutoff at the optical frequencies.

The difference in the maximum possible frequencies of photons the IC of which makes the main contribution to the X-ray radiation of the near and far knots leads to different frequencies of the high-energy cutoff of the spectra for the same values of $\Gamma_{\text{max}}$ in the knots. 
When modeling the spectrum of scattered radiation, we adopted $\Gamma_{\text{max}}=10^5$. 
It can be seen from Fig.~\ref{fig:3c273gamma} that the model spectrum of the knots~A and B2 at frequencies corresponding to the photon energy of 1–10 GeV is approximately equal to the upper bound of the expected flux from the kpc-scale jet of 3C~273,
established by \citet{MeyGeor14}. If the value of $\Gamma_{\text{max}}$ for these knots is taken smaller or a detailed study of the CS spectrum shows lower values of the maximum frequency of the 2nd part, then the high-energy cutoff of the X-ray jet spectrum will occur at lower frequencies.
Then, the gamma-ray flux will be significantly lower than the
upper bound of the expected gamma-ray flux from the kpc-scale jet determined from the \textit{Fermi}-LAT data \citep{MeyGeor14}. Therefore, based on the available data, the assumption about IC/CS as a X-ray emission mechanism for the nearest to the quasar knots~A and B2 cannot be rejected.

\subsection{PKS~1510---089}

In the X-ray range in the quasar PKS~1510$-$089 jet, knots A, B, and C (in order of increasing distance from CS) are detected; they correspond to the knots in the radio range, except the last, most distant knot \citep{Samb04}. 
The X-ray intensity of the knots decreases with the distance from CS, which suggests that IC/CS is the X-ray emission mechanism for all the knots. 
Let us consider the scattered radiation and its spectrum for each knot separately.

The X-ray spectral index of the knot A is $\alpha_{\text{X}}\approx\alpha_1\neq\alpha_{\text{R}}$; therefore, the main contribution to the X-ray flux from the knot is made by IC of the 1st part of the CS spectrum under the limitation from the electron spectrum. 
Since $\alpha_{\text{X}}=0.09\pm0.43$ \citep{Samb04} and $\alpha_{\text{R}}=0.6$ (for the entire jet) \citep{ODea1988}, electrons with $\Gamma_{\text{min}}$ are scattered most efficiently. This process produces photons with frequencies ranging from $k_{\text{IC}}\, \omega_{\text{min, j}}\Gamma_{\text{min}}^2$ to $\omega_{\text{X, br1}}=k_{\text{IC}}\,\omega_{\text{0, j}}\Gamma^2_{\text{min}}$.
If we consider the spectrum of scattered radiation in the range corresponding to the photon energy of 0.5$-$8~keV to be power-law, then we have $1.4\cdot10^3<\Gamma_{\text{min}}<1.7\cdot10^4$.
At the frequency $k_{\text{IC}}\,\omega_{\text{min, j}}\Gamma_{\text{min}}^2$, the scattered radiation spectrum is cutoff.
At frequencies from  $\omega_{\text{X, br1}}$ to $\omega_{\text{X, br2}}=k_{\text{IC}}\omega_{\text{0, j}}\Gamma_{\text{max}}^2$, radiation is formed due to IC of photons with a frequency $\omega_{\text{0, j}}$ on electrons of a power-law energy distribution.
The spectral index of this radiation is $\left(\gamma-1 \right)/2$. 
The IC of the 2nd part of the SC spectrum by electrons with $\Gamma_{\text{max}}$ (since $\alpha_{\text{R}}<\alpha_2$) produces radiation at frequencies from $\omega_{\text{X, br2}}$ to $k_{\text{IC}}\,\omega_{\text{max, j}}\Gamma_{\text{max}}^2$, above which the spectrum of scattered radiation is cutoff.

The X-ray spectral index of the knot~B  $\alpha_{\text{X}}=0.81\pm0.62$ \citep{Samb04}, was determined with low accuracy. Therefore, it can be equal to both $\alpha_2$ and $\alpha_{\text{R}}$. The large error in $\alpha_{\text{X}}$ can partly be caused by the break in the X-ray emission spectrum. If $\alpha_{\text{X}}=\alpha_2$, then the X-ray radiation of the knot~B is formed due to the scattering of the 2nd part of the CS spectrum on electrons with $\Gamma_{\text{max}}$.
In this case, the minimum frequency of the scattered radiation, according to formula~(\ref{eq:osv}), must be $k_{\text{IC}}\,\omega_{\text{0, j}}\Gamma_{\text{max}}^2\leq\omega_{\text{X}}$.
This condition yields $\Gamma_{\text{max}}\leq5\cdot10^2$. 
This value seems to be too small even if we assume that it corresponds to the Lorentz factor of the break in the electron energy spectrum. Then, possibly, $\alpha_{\text{X}}=\alpha_{\text{R}}$ and X-ray radiation is formed due to IC with the limitation imposed by the photon spectrum. Considering IC of the 1st and 2nd parts by electrons with a power-law energy distribution, we found that the scattering of photons with a frequency $\omega_{\text{0, j}}$ makes the main contribution to the observed X-ray radiation.

\begin{figure}
\center{ \includegraphics[scale=0.5]{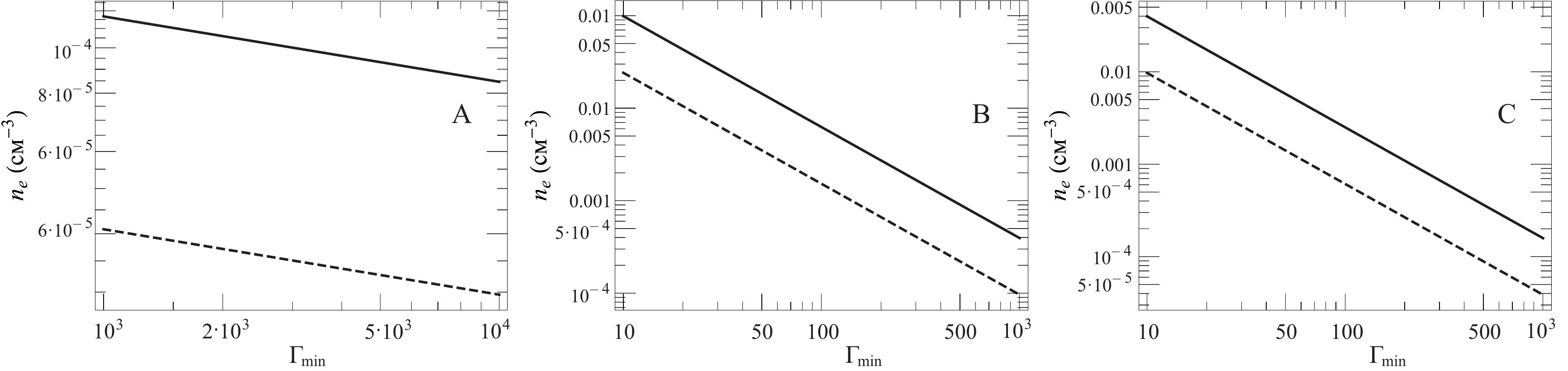}  }  
 \caption{
Number density of emitting electrons in the knots~A, B, and C of the PKS~1510$-$089 jet, which have detectable X-ray radiation.
The solid line corresponds to electron number density for $\theta_{\rm kpc}=24^\circ$ and dashed line --- for $\theta_{\text{kpc}}=34^\circ$.
 }
 \label{fig:1510ne}
\end{figure}

\begin{figure}
\center{ \includegraphics[scale=0.65]{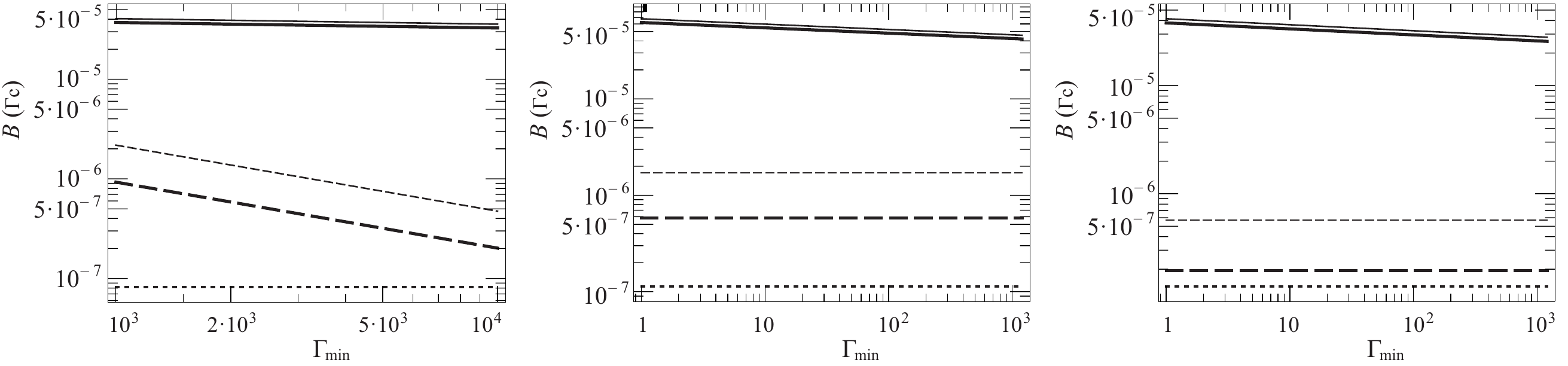}  }  
 \caption{
 Magnetic field in three knots of the PKS~1510$-$089 jet, which have detected X-ray radiation: (solid lines) the magnetic
field strength under the energy equipartition condition; (dashed lines) the magnetic field strength obtained from a comparison
of radio and X-ray fluxes within IC/CS for $\theta_{\text{kpc}}=24^\circ$ (thick dashed lines) and $\theta_{\text{kpc}}=34^\circ$. (thin dashed lines); (dotted lines)
the magnetic field determined under the assumption of IC/CMB. All values of the magnetic field are shown for $\delta_{\text{kpc}}=1$.
 }
 \label{fig:1510H}
\end{figure}

The radio and X-ray spectral indices of the knot~C are approximately equal; therefore, X-ray radiation is formed due to IC with a limitation imposed by the photon spectrum, as well as in the knot~B. The shape of the scattered radiation spectrum of the knots~B and C is similar to the case of the knot~A, taking into
account that  $\omega_{\text{X, br1}}$ and $\omega_{\text{X, br2}}$ depend on the parameters of the electron energy distribution, which may be different in these two knots. Taking into account the specificities of IC/CS in the knots of the PKS~1510$-$089 jet, we estimated the density of emitting electrons
and the magnetic field strength at the knots (Table~\ref{tab:par1510}, Figs.~\ref{fig:1510ne} and \ref{fig:1510H}).

\begin{table*}
 \caption{
Physical parameters for the knots~A, B, and C of the PKS~1510$-$089 quasar, jet, determined under the assumption of IC/CS. The columns represent: (1) the knot; (2) $\theta_{\text{kpc}}$ used in calculations; (3), (4) the electron density for $\Gamma_{\text{min}}=10$ and $\Gamma_{\text{min}}=100$, respectively; (5) the magnetic field determined from the ratio of the radio and X-ray fluxes, in the calculations for the knot~A, $\Gamma_{\text{min}}=10$ was used, and the values for the rest knots do not depend on $\Gamma_{\text{min}}$; (6) the magnetic field determined from the radio emission of the knot,
necessary to fulfil the energy equipartition condition for $\Gamma_{\text{min}}=10$; (7) the magnetic field determined under the assumption of IC/CMB for $\delta_{\text{kpc}}=1$ and the values given for the knot A were obtained for $\Gamma_{\text{br}}=10^4$; (8) the boundaries of the power-law electron distribution.
 } 
 \label{tab:par1510} 
 \medskip
  \begin{flushleft}
 \begin{tabular}
 {|c|c|c|c|c|c|c|c|}
  \hline
Knot & $\theta_{\rm kpc}$, deg & $n_{e10}$, cm$^{-3}$ & $n_{e100}$, cm$^{-3}$ & $B_{\perp}$, G & $B_{\text{eq}}$, G & $B_{\perp, \rm CMB}$, G & $e^{-}$ spectrum \\
  \hline
(1) & (2) & (3) & (4) & (5) & (6) & (7) & (8) \\
\hline
 \multirow{2}{*}{A} &  24 & 0.34 & 0.02 & $2.0 \cdot 10^{-7}$ & $4.8 \cdot 10^{-5}$ & \multirow{2}{*}{$8.2 \cdot 10^{-8}$} & \raisebox{-0.7ex}[0cm][0cm]{$1.4\cdot10^3<\Gamma_{\text{min}}<1.7\cdot10^4$} \\
 \cline{2-6}
            & 34  & 0.12  & $7 \cdot10^{-3}$  & $4.7 \cdot 10^{-7}$  & $5.2 \cdot 10^{-5}$  &  & \raisebox{0.5ex}[0cm][0cm]{$\Gamma_{\text{max}}<10^6$}\\
\hline
 \multirow{2}{*}{B} &  24 & 0.10 & $6\cdot10^{-3}$ & $5.8 \cdot 10^{-7}$ & $5.5 \cdot 10^{-5}$ & \multirow{2}{*}{$1.1 \cdot 10^{-7}$} & \multirow{2}{*}{\raisebox{-1.8ex}[0cm][0cm]{$\Gamma_{\text{min}}\leq355$}} \\
 \cline{2-6}
            & 34  & 0.02  & $2\cdot10^{-3}$  & $1.7 \cdot 10^{-6}$  & $4.8 \cdot 10^{-5}$  & &  \\
\cline{1-7}
 \multirow{2}{*}{C} &  24 & 0.04 & $3\cdot10^{-3}$ & $1.9 \cdot 10^{-7}$ & $3.4 \cdot 10^{-5}$ & \multirow{2}{*}{$1.4 \cdot 10^{-7}$} & \multirow{2}{*}{\raisebox{1.8ex}[0cm][0cm]{$\Gamma_{\text{max}}<5\cdot10^4$}} \\
 \cline{2-6}
            & 34  & 0.01  & $6\cdot10^{-4}$  & $5.7 \cdot 10^{-7}$  & $3.7 \cdot 10^{-5}$  & & \\
\hline
  \end{tabular}
   \end{flushleft}
\end{table*}

\begin{figure}
    \centering
    \includegraphics[scale=0.7]{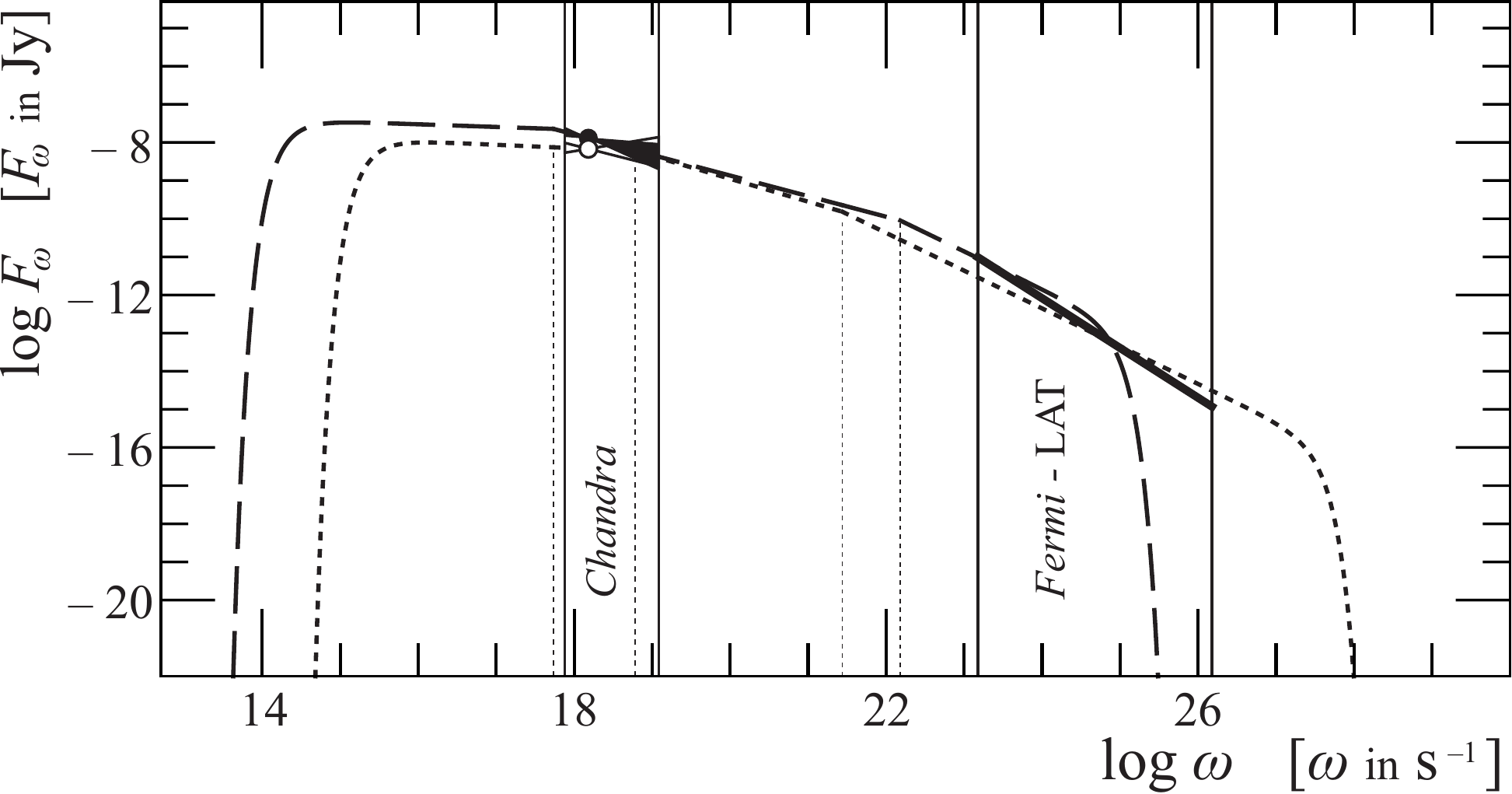}
    \caption{
    Spectra of scattered radiation from the knots of the kpc-scale jet of the quasar PKS~1510$-$089. The spectrum of the knot~A
(short dashes) was plotted with the parameters $\Gamma_{\text{min}}=10^3$ and $\Gamma_{\text{max}}=10^6$. 
The total spectrum of the knots B and C (long dashes) is plotted with the parameters $\Gamma_{\text{min}}=300$ and $\Gamma_{\text{max}}=5\cdot10^4$. 
The observed fluxes from the knot A (white dots) and the
total flux from knots B and C (black dots) are shown with the spectral index measurement errors, which are marked by the triangles filled with the corresponding color. The Chandra and Fermi-LAT operating ranges are represented by solid vertical lines. Bold solid line displays the minimum flux considered as the upper bound of the constant gamma-ray flux from the kpc-scale jet, with the energy spectral index $\alpha_{\text{X}}=1.3$.
Dashed vertical lines mark the frequencies of the break in the spectrum of scattered radiation. With the given parameters, these frequency for the knot A are $6\cdot10^{18}$~s$^{-1}$ and $2.7\cdot10^{21}$~s$^{-1}$, for the knot B and C --- $5.4\cdot10^{17}$~s$^{-1}$ and $1.5\cdot10^{22}$~s$^{-1}$.}
    \label{fig:1510gamma}
\end{figure}

The spectrum of scattered radiation of the knot~A and the total emission spectrum of the knots~B and~C are shown in Fig.~\ref{fig:1510gamma}. 
The values of $\Gamma_{\text{min}}$ and $\Gamma_{\text{max}}$ were
chosen so as, firstly, to exclude the production by
IC/CS of an optical flux at a level sufficient for detection;
secondly, the breaks in the spectrum of scattered
radiation of the knots should not occur in the middle
of the observed X-ray range; thirdly, the integral theoretically
expected gamma-ray flux from the knots~A,
B, and C should be consistent with the upper bound of
the constant flux. This upper bound can be taken
equal to the minimal flux from the entire object
observed by \textit{Fermi}-LAT. We adopted this flux at the
level of $\approx5\cdot10^{-11}$~erg~cm$^{-2}$~s$^{-1}$ in the range $0.1-100$~GeV based on both 8-year data with weekly averaging and the data from the 3FGL catalog~\cite{Acero15} with
one month averaging over the first four years of \textit{Fermi}-LAT observations. 
With $\Gamma_{\text{max}}=10^6$ for the knot~A and $\Gamma_{\text{max}}=10^4$ for the knots~B and C, the total
flux from the knots in the \textit{Fermi}-LAT operating range
is $3\cdot10^{-11}$~erg~cm$^{-2}$~s$^{-1}$. 
This value can be lower if $\Gamma_{\text{max}}$ is taken smaller, especially for the knots~B and C.
Thus, from the limitation on the gamma-ray flux, an
estimate of the maximum energy of ultrarelativistic
electrons in the knots can be obtained.

\subsection{PKS~1045---188}

The quasar PKS~1045$-$188 jet in the X-ray range is
detected up to the scales of $\approx8^{\prime\prime}$ \citep{HoganList11, Massaro2011} and is not detected in the jet region after the bend observed on
the VLA radio maps at a frequency of 1.4~GHz \citep{KharbLC10}.
For this object, there are no published data on the
spectral index and spectral flux from the knots in the
X-ray range. We found these parameters by processing
the \textit{Chandra} observations no. 15037 using the CIAO~4.10 program with the CALDB~4.7.9 calibration files.
The flux and spectral index were determined for two
knots from the parts bounded by circles with a diameter
of $1.13^{\prime\prime}$, located at distances of $\approx4^{\prime\prime}$ and $7^{\prime\prime}$ from the
central source. The obtained values are given in
Table~\ref{tab:1045Xinfo}. 

\begin{table*}
 \caption{Observation data for the PKS~1045$-$188 jet knots in the X-ray and radio ranges. The columns represent (1) the nomenclature of knots used in this article, (2) the distance from CS, (3) integral flux in the range 0.5$-$7~keV, (4) the flux at a frequency corresponding to a photon energy of 1 keV, (5) the X-ray spectral index, and (6) the flux density at a frequency of 1.43~GHz \citep{Massaro2011}.
 } 
 \label{tab:1045Xinfo} 
 \medskip
 \begin{tabular}
 {|c|c|c|c|c|c|c|c|c|c|c|}
  \hline
 Knot & $R$, $^{\prime\prime}$ & $F_{0.5-7}$, erg~cm$^{-2}$~s$^{-1}$ & $F_1$, erg~cm$^{-2}$~s$^{-1}$~Hz$^{-1}$ & $\alpha_{\rm X}$ & $F_{\rm R}$, mJy \\
  \hline
 (1) & (2) & (3) & (4) & (5) & (6) \\
  \hline
  A & 4 & $1.07\cdot 10^{-14}$ & $1.46\cdot10^{-32}$ & $0.87\pm0.38$ & 24.4 \\
  B & 7 & $4.76\cdot10^{-15}$ & $6.06\cdot10^{-33}$ & $0.78\pm1.37$ &78.3 \\
  \hline
  \end{tabular}
  \end{table*}

Also, there are not published data on the radio spectral
index for the kpc-scale jet, but, assuming that the main contribution to the
low-frequency ($\omega\leq3\cdot10^9$~s$^{-1}$) part of the CS spectrum
(Fig.~\ref{fig:CSspectra}) is made by the radiation generated in the
kpc-scale jet, we adopt for further calculations $\alpha_{\text{X}}=\alpha_3\approx0.6$.
For example, a similar situation takes
place for 3C~273 and PKS~0637$-$752, because the
spectral indices of these kpc-scale jets correspond to
the spectral indices obtained by linear approximation
of the low-frequency CS spectra (part~3 in Fig.~\ref{fig:CSspectra}).

Within the measurement errors, for the knot~A, we
have $\alpha_{\text{X}}\approx\alpha_{\text{R}}$ and $\alpha_{\text{X}}\approx\alpha_2$. 
In the latter case, the X-ray radiation is formed due to IC of the 2nd part of
the CS spectrum by electrons with $\Gamma_{\text{max}}$ (see formula
(\ref{eq:FwxElec}) and the explanations to it).
Then, using expression (\ref{eq:osv}), from the absence of a break in the spectrum of scattered radiation in the frequency range corresponding
to photon energies of 0.5$-$7~keV, we find the limitations $200\leq\Gamma_{\text{max}}\leq900$. These values are small even
if we assume that they correspond to the Lorentz factor at which a break in the power-law electron energy distribution occurs.

\begin{figure}
    \centering
    \includegraphics[scale=0.6]{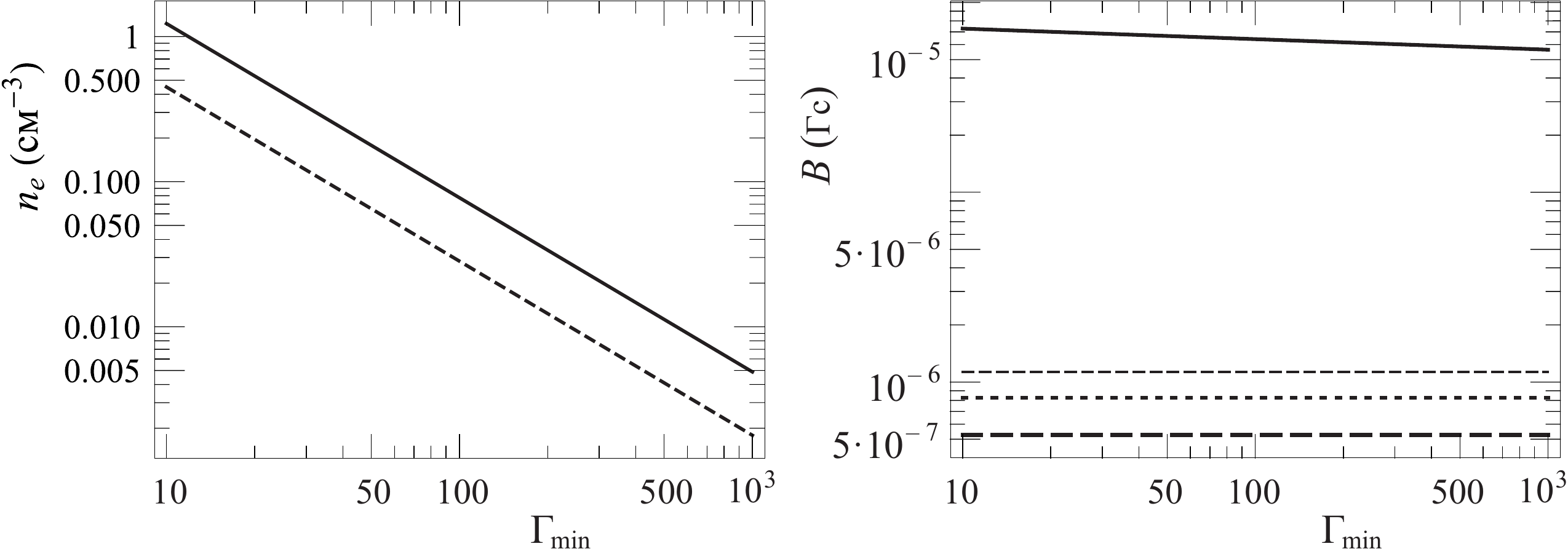}
    \caption{Number density of emitting particles (a) and magnetic field strength (b) in the knot~A of the quasar PKS~1045$-$188 jet. Left panel: $\theta_{\text{kpc}}=34^\circ$ (solid line) and $\theta_{\text{kpc}}=44^\circ$ (dashed line). Right panel: The magnetic field found from a comparison of the observed radio and X-ray fluxes within the IC/CS for $\theta_\text{kpc}=34^\circ$ (thick dashed line) and for  $\theta_{\text{kpc}}=44^\circ$ (thin dashed line). The magnetic field under the energy equipartition condition for $\theta_{\text{kpc}}=34^\circ$ is denoted by solid line. The magnetic field obtained within IC/CMB for $\delta_{\text{kpc}}=1$ is shown by dotted line. }
    \label{fig:1045neH}
\end{figure}

In the case of $\alpha_{\text{X}}\approx\alpha_{\text{R}}$, the X-ray radiation of the knot~A, as well as the knot~B, is formed due to the scattering
of photons with a frequency $\omega_{\text{0, j}}$ on electrons
with the power-law energy distribution. The estimated
values of the number density of emitting electrons and the
magnetic field strength in the knots of the PKS~1045$-$188 jet are consistent with similar values obtained by us for the jets of other sources and are presented in Table~\ref{tab:par1045} and Fig.~\ref{fig:1045neH}.

\begin{table*}
 \caption{
Physical parameters for the knots of the quasar PKS~1045$-$188 jet, determined under the assumption of IC/CS.  The columns represent: (1) the knot; (2) $\theta_{\rm kpc}$ used in calculations; (3) and (4) the electron number densities for $\Gamma_{\rm min}=10$ and $\Gamma_{\rm min}=100$, respectively; (5) the magnetic field determined from the ratio of radio and X-ray fluxes; (6) the magnetic field determined from the radio emission of the knot, which is necessary to fulfil the energy equipartition condition for $\Gamma_{\rm min}=10$; (7) the magnetic field determined under the assumption of IC/CMB for $\delta_{\rm kpc}=1$; and (8) the boundaries of the power-law distribution of electrons.
} 
 \label{tab:par1045} 
 \medskip
 \begin{tabular}
 {|c|c|c|c|c|c|c|c|}
  \hline
Knot & $\theta_{\rm kpc}, ^\circ$ & $n_{e10}$, cm$^{-3}$ & $n_{e100}$, cm$^{-3}$ & $B_{\perp}$, G & $B_\text{eq}$, G & $B_{\perp, \rm CMB}$, G & $e^{-}$ spectr\\
  \hline
(1) & (2) & (3) & (4) & (5) & (6) & (7) & (8) \\
\hline
 \multirow{2}{*}{A} &  34 & 1.23 & 0.08 & $5.3 \cdot 10^{-7}$ & $7.3 \cdot 10^{-5}$ & \multirow{2}{*}{$8.3 \cdot 10^{-7}$} & \multirow{2}{*}{\raisebox{-1.8ex}[0cm][0cm]{$\Gamma_\text{min}<1.7\cdot10^3$}} \\
 \cline{2-6}
            & 44  & 0.45  & 0.03  & $1.1 \cdot 10^{-6}$  & $7.7 \cdot 10^{-5}$  & & \\
\cline{1-7}
 \multirow{2}{*}{B} &  34 & 1.57 & 0.99 & $5.3 \cdot 10^{-6}$ & $1.0 \cdot 10^{-4}$ & \multirow{2}{*}{$3.0 \cdot 10^{-6}$} & \multirow{2}{*}{\raisebox{1.8ex}[0cm][0cm]{$6.2\cdot10^3<\Gamma_\text{max}\leq5\cdot10^5$}}\\
 \cline{2-6}
            & 44  & 1.26  & 0.08  & $1.5 \cdot 10^{-5}$  & $1.1 \cdot 10^{-4}$  & & \\
\hline
 
  \end{tabular}
\end{table*}

\begin{figure}
    \centering
    \includegraphics[scale=0.7]{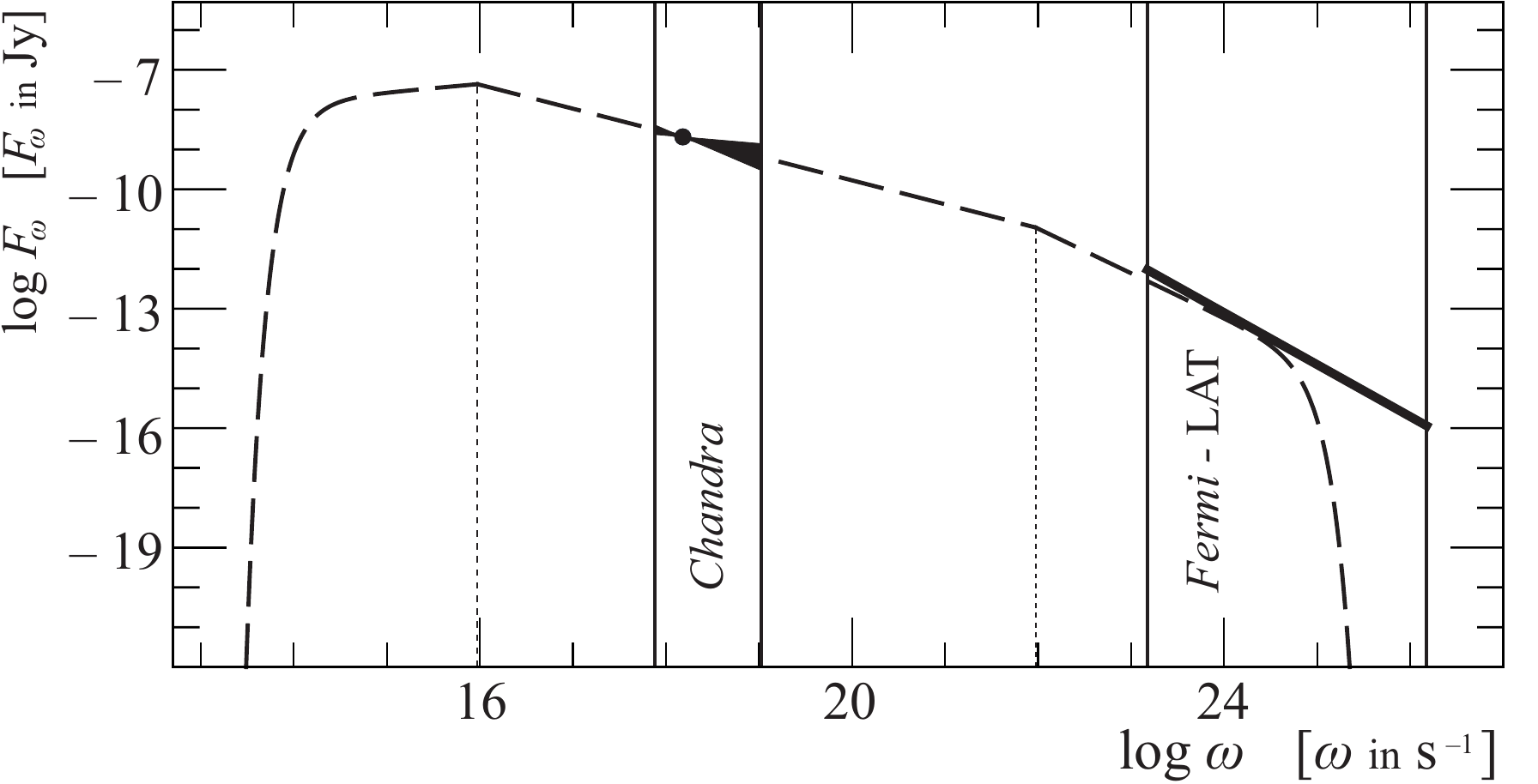}
    \caption{
    Spectrum of scattered radiation from the knots~A and B of the kpc-scale jet of the quasar PKS~1045$-$188 (dashed line).
    The spectrum was plotted for the parameters $\Gamma_{\text{min}}=100$, $\Gamma_{\text{max}}=5\cdot10^5$. 
    The total observed flux from the knots~A and B are denoted by black dot and measurement errors of the spectral index are marks by filled triangles. The solid vertical lines denote the \textit{Chandra} and \textit{Fermi}-LAT operating ranges.
    The thick solid line indicates the minimum flux, which is considered as the upper bound of the constant gamma-ray flux from the kpc-scale jet for the energy spectral index $\alpha_\text{X}=1.5$. Dotted vertical lines show the frequencies of the break in the spectrum of scattered radiation caused by the transition from the limitation imposed by the electron spectrum to the limitation imposed by the photon spectrum
(low-frequency break) and from the limitation imposed by the photon spectrum to limitation imposed by the electron spectrum
(high-frequency break).
    }
    \label{fig:1045gamma}
\end{figure}

Figure~\ref{fig:1045gamma} shows the simulated spectrum of scattered
radiation from the knots~A and B. The radiation
of the low-frequency part of this spectrum is formed
due to IC of the 1st part of the CS spectrum by electrons
with $\Gamma_{\text{min}}$. The spectral index of this radiation is $\alpha_{\text{X}}=\alpha_1$. At the middle frequencies, including the
\textit{Chandra} operating range, radiation is formed due to
IC of photons with $\omega_{\text{0, j}}$ on electrons of the power-law
energy spectrum. In this case $\alpha_{\text{X}}=\alpha_{\text{R}}$.
High-frequency radiation is formed due to IC of the 2nd part
of the CS spectrum by electrons with $\Gamma_{\text{max}}$. In this
case $\alpha_{\text{X}}=\alpha_2$. Figure~\ref{fig:1045gamma} also shows the upper bound of the gamma-ray flux from the kpc-scale jet, which
we estimated based on the following. The radio source
PKS~1045$-$188 is not positionally associated with any of the detected gamma-ray sources in the sky within an error of $95\%$, according to \textit{Fermi}-LAT data \citep{Abdo10}.
Moreover, it is not in the plane of the Galaxy ($b=35^\circ$).
Therefore, the sensitivity of the telescope was set as the upper bound of the flux
from this object, according to the typical photon index of quasars $\alpha_\gamma=1.5$~\citep{Abdo10}, which is about $4.6\cdot10^{-12}$~erg~sm$^{-2}$~s$^{-1}$ at energies above 100~MeV \citep{Abdo10}.
The integral flux from the knots~A and B in the range 0.1$-$100~MeV depends on $\Gamma_{\text{max}}$, since this quantity determines the position of the high-frequency cutoff in the spectrum of scattered
radiation. At $\Gamma_{\text{max}}=5\cdot10^5$, the integral theoretical
flux is equal to $2\cdot10^{-8}$~erg~sm$^{-2}$~s$^{-1}$, which is smaller
than the estimate for the flux from the kpc-scale jet,
while at $\Gamma_{\text{max}}=10^6$, the theoretically expected flux on
the level of $8\cdot10^{-8}$~erg~sm$^{-2}$~s$^{-1}$ exceeds the upper estimate
for the flux from the kpc-scale jet.

\section{DISCUSSION AND CONCLUSIONS}

We have considered the inverse Compton scattering of radiation from a pc-scale jet on ultrarelativistic electrons of a kpc-scale jet as a X-ray emission mechanism for the jets of the quasars PKS~0637$-$752, 3C~273, PKS~1510$-$089, and PKS~1045$-$188.
For all the sources, we have obtained adequate estimates for the density of emitting
particles ($\sim1-10^{-3}$~sm$^{-3}$) and the magnetic field strength ($\sim10^{-5}-10^{-6}$~G for $\delta_{\text{kpc}}=1$).
If we take into account the moderate relativistic motion of the kpc-scale
jets, then the magnetic field in the reference frame of the kpc-scale jet will be several times stronger.
However, for all considered sources, as
well as for PKS~1127$-$145~\citep{BP19}, the condition of equipartition
is not fulfilled: the energy density of the particles
is greater than the energy density of the magnetic field.
This inequality becomes even stronger if we
assume the presence of other particles in the kpc-scale
jet additionally to the electron–positron plasma. The possibility
of such a deviation from equipartition for kpc-scale jets was noted by \citet{Mehta09}.
An extremely high brightness temperature, exceeding the maximum permissible
value under the equipartition condition excluding the Compton catastrophe, was detected by the ground–space radio interferometer \textit{RadioAstron} for
several active nuclei \citep{Gomez16, Kutkin18, Pilipenko18, Vega19, Kravchenko20}, including 3C~273 \citep{Kovalev16}.
This result can be explained either by the large ($\sim100$)
Doppler factor of the pc-scale jets, or by the fact that
the energy density of the emitting particles is higher
than the energy density of the magnetic field even in
the pc-scale. Therefore, in our opinion, the energy equipartition condition cannot be dominant in determining the X-ray emission mechanism, acting in the kpc-scale jets of quasars.

Unlike other models, IC/CS explains without
additional assumptions the observed decrease in the
X-ray intensity of the knots with the distance from the
CS. For the outer knots of the 3C~273 jet, the X-ray
intensity has a small and approximately constant value.
This fact \citet{MBK10} interpreted by that the X-rays from these knots are due to IC/CMB. From the comparison of the flux densities formed by IC/CS and IC/CMB,
an interval of the kpc-scale jet angle with the line of sight of $25-26^\circ$
has been found. For the kpc-scale jets of the quasars PKS~0637$-$752, PKS~1045$-$188, and PKS~1510$-$089, the observed X-ray radiation from the knots of which is formed only due to IC/CS, a lower bound of the angle between the jet and the line of sight ($\sim25^\circ$) has been obtained. 
The difference in the angles between the line of sight and pc-scale jets, estimated
from the apparent superluminal motion of their
features \citep{Lister19, EdwardsPT06}, and the angles between the line of
sight and kpc-scale jets can be explained by the deceleration of jets between the pc- and kpc-scales. It is possible that this deceleration already manifests itself at distances of about 100~pc from the core \citep{Homan15,Piner12}.
In this case, the value of the Lorentz factor change is $\dot\Gamma/\Gamma\approx10^{-3}-10^{-2}$~\cite{Homan15}, and, if that changes insignificantly, then over a time of hundreds to thousands of
years in the reference frame of the source, the jet can
slow down from the Lorentz factor from 10 to 1.
We have found that the kpc-scale jets make up an angle of $\geq25^\circ$ with the line of sight and have an average velocity of $(0.6-0.95)c$.
The velocities are consistent with other
independent estimates of the velocities of the kpc-scale
jets of active galaxies \citep{WA97, ArLon04, MullinH09}.

IC/CS predicts the existence of breaks in the spectrum
of scattered radiation. These breaks are caused by
the transition from the limitation imposed by the electron
spectrum to the limitation imposed by the photon
spectrum and vice versa. The detection of breaks in
the X-ray spectrum of the knots of the kpc-scale jets
both will prove the action of IC/CS, and will make it
possible to determine the parameters of the electron
spectrum.

In the frame work of IC/CS, the gamma-ray
flux was simulated for each source. The flux, obtained in the frequency
range corresponding to the photon energies of 0.1$-$100~GeV depends on the choice of the maximum Lorentz factor of electrons. For adequate values of $\Gamma_{\text{max}}$, this flux turns out to be lower than the constant flux, obtained from the \textit{Fermi}-LAT data. On
the assumption that this constant flux is generated in a
kpc-scale jet, IC/CS does not contradict any available
observational data.

\acknowledgments
This work was partly supported by the Russian Foundation
for Basic Research, project no.~18-32-00824. Within
this project, approximations of the CS spectra were obtained for all considered objects, the density of emitting electrons and the magnetic field strength at the
knots of the kpc-scale jets were estimated, the angles with
the line of sight and the velocities of the kpc-scale jets were
determined, and the spectrum of radiation due to IC/CS
was simulated.

\bibliography{kpcjets}
\bibliographystyle{aasjournal}

\end{document}